\documentclass[preprint,aps,superscriptaddress]{revtex4-1} 

\usepackage{epsf}
\usepackage{epsfig}
\usepackage{color}
\usepackage{graphicx}
\usepackage{amsmath}
\usepackage{bbm}

\begin{document}

\title
{Non-Coplanar Model States in Quantum Magnetism Applications of the High-Order Coupled Cluster Method}

\author{D. J. J. Farnell}
\email[email: ]{FarnellD@cardiff.ac.uk}
\affiliation{School of Dentistry, Cardiff University, Cardiff CF14 4XY, Wales, UK}

\author{R. F. Bishop}
\email[email: ]{raymond.bishop@manchester.ac.uk}
\affiliation{School of Physics and Astronomy, Schuster Building, 
The University of Manchester, Manchester M13 9PL, UK}
\affiliation{School of Physics and Astronomy, University of Minnesota, 116 Church Street SE, Minneapolis, MN 55455, USA}

\author{J. Richter}
\email[email: ]{Johannes.Richter@Physik.Uni-Magdeburg.DE}
\affiliation{Institut f\"ur Physik, Otto-von-Guericke-Universit\"at Magdeburg,
D-39016 Magdeburg, Germany}
\affiliation{Max Planck Institute for the Physics of Complex Systems, D-01187 Dresden, Germany}

\date{\today}

%\begin{history}
%\received{Day Month Year}
%\revised{Day Month Year}
%\accepted{(Day Month Year)}
%\comby{(xxxxxxxxxx)}
%\end{history}

\begin{abstract}Coplanar model states for applications of the coupled cluster method 
(CCM) to problems in quantum magnetism are those in which all spins lie in a plane, 
whereas three-dimensional (3D) model states are, by contrast, 
non-coplanar ones in which all the spins do not lie in any single plane.  A crucial first step in 
applying the CCM to any such lattice quantum spin system is to perform a passive rotation of 
the local spin axes so that all spins in the model state appear mathematically to 
point in the same (say, downwards $z$-)direction. Whereas this process leads to terms with 
only real coefficients in the rotated Hamiltonian for coplanar model states, an additional
complication arises for 3D model states where the corresponding coefficients
can become complex-valued.  We show here for the first time how high-order 
implementations of the CCM can be performed for such Hamiltonians.  
We explain in detail why the extension of the computational 
implementation of the CCM when going from coplanar to 3D model states is a non-trivial task 
that has not hitherto been undertaken.
To illustrate these new developments, we present results for three cases: 
(a) the spin-half
one-dimensional Ising ferromagnet in an applied
transverse magnetic field (as an exactly solvable test model to use as a yardstick 
for the viability and accuracy of our new methodology); (b) the spin-half
triangular-lattice Heisenberg antiferromagnet 
in the presence of an external magnetic  field; and (c) the spin-$S$
triangular-lattice {\it XXZ} antiferromagnet 
in the presence of an external magnetic field, for the cases $\frac{1}{2} \leq S \leq5 $. 
For 3D model
states the sets of algebraic CCM equations for the ket- and bra-state 
correlation coefficients become complex-valued, but  ground-state expectation 
values of all physical observables are manifestly real numbers, as required, and as we explicitly demonstrate in all three applications.  Indeed, excellent correspondence is 
seen with the results of other methods, where they exist, for these systems. 
In particular, our CCM results demonstrate explicitly that 
coplanar ordering is favoured over non-coplanar 
ordering for the triangular-lattice spin-half Heisenberg antiferromagnet at all values of the applied 
external magnetic field, whereas for the anisotropic {\it XXZ} model 
non-coplanar ordering can be favoured in some regions of the parameter space.
Specifically, we present a precise determination of the boundary (i.e.,  the critical 
value of the {\it XXZ} anisotropy  parameter $\Delta$) between 
a 3D ground state and a coplanar ground 
state for the {\it XXZ} model for values for the external magnetic field 
near to saturation, for values of the spin quantum number $S \leq 5$. Although the CCM calculations are computationally intensive for this frustrated model, especially for high 
spin quantum numbers, our accurate new results certainly improve our understanding of it. 
\end{abstract}


\maketitle

\section{Introduction}

The coupled cluster method (CCM)
\cite{refc1,refc2,refc3,refc4,refc5,refc6,refc7,refc8,refc9,RFB_1998} 
is a powerful method of quantum many-body theory that has long
been used to study strongly interacting and highly frustrated 
quantum spin systems with great success 
\cite{ccm1,ccm2,ccm3,ccm4,ccm5,ccm6,ccm7,ccm7a,ccm8,CCM_shastry,ccm9,ccm10,ccm11,
ccm12,ccm13,ccm14,kago1,ccm15,ccm16,ccm17,ccm18,kago2,kago3,ccm19,bishop2017,
bishop2018}.  
The introduction of the ``high-order'' CCM \cite{ccm5,ccm6,ccm7} for these 
systems has led to a step-change in its accuracy.  The high-order CCM employs 
very high orders of approximation schemes, for which the equations
that determine all multispin correlations retained at 
any given level are both derived and subsequently solved by using 
massively parallel computational tools \cite{ccm5,ccm6,ccm7,code}.
The CCM is now fully competitive with  the best of other approximate methods,
especially for systems of $N$ spins on the sites of
a lattice in two (see, e.g., Refs. \cite{kago1,ccm18,
kago2,bishop2018} and references cited therein) or three 
(see, e.g., Refs. \cite{kago3,ccm19} and references cited therein) 
spatial dimensions.  Unlike several other approximate quantum many-body
methods that are limited in their range of applicability by frustration
(i.e., where bonds in the Hamiltonian compete against each other to 
achieve energy minimisation), the CCM 
has been applied previously even to highly frustrated and strongly correlated quantum spin systems with much success. Thus, recently it has been demonstrated, for example,
that the high accuracy needed to investigate the quantum ground-state selection
of competing states of the kagome antiferromagnet is provided by high-order
CCM calculations \cite{kago1,kago2,kago3}. 
Another advantage of the high-order CCM  is 
that it is very flexible. For example, both in principle and 
in practice, the CCM technique can treat essentially all Hamiltonians 
containing either single spin operators and/or products of two spin operators, on
any crystallographic lattice, and for any spin quantum number $S$.
We note too that, unlike most alternative techniques, the CCM
can be applied from the outset in the thermodynamic limit, $N \to \infty$, at 
{\em every} level of approximate implementation, thereby
obviating the need for any finite-size scaling of the results.

In all practical implementations of the CCM the many-body
correlations present in the exact (ground or excited) state of the system under investigation
are expressed with respect to a suitable model (or reference) state, 
as we explain in more detail below in Sec.~\ref{ccm}.
Coplanar model spin states used in the CCM are those states in which all spins 
lie in a plane, whereas three-dimensional (3D) model spin states are non-coplanar 
states in which the spins do not lie in any one plane. We remark that, 
until now, only coplanar model spin states have been used in all prior 
CCM calculations in the field of quantum magnetism 
for reasons that we now explain.

Thus, an important ingredient used in all practical applications of the CCM 
to spin-lattice systems
\cite{ccm1,ccm2,ccm3,ccm4,ccm5,ccm6,ccm7,ccm7a,ccm8,CCM_shastry,ccm9,ccm10,
ccm11,ccm12,ccm13,ccm14,kago1,ccm15,ccm16,ccm17,ccm18,kago2,kago3,
ccm19,bishop2017,bishop2018} 
is to rotate the local spin axes of all spins in the model state such that they appear 
(mathematically only) to be point in the ``downwards'' $z$-direction. One 
can always choose a set of rotations that leads to terms in the 
Hamiltonian that contain only real-valued coefficients, 
with respect to the new set of local spin axes, for the coplanar model 
states.  By contrast,  three-dimensional (3D) (non-coplanar) model states 
inevitably lead to  terms in the new Hamiltonian after rotation of the local spin axes
that contain complex-valued coefficients.  These cases 
are more difficult to treat both analytically and computationally. Of course, all
macroscopic physical parameters calculated within the CCM, such as the ground-state energy 
and magnetic order parameter, still have to be real numbers because the transformations of 
local spin axes are unitary and the resulting Hamiltonian is still Hermitian. 
Nevertheless, the intervening multispin correlation coefficients
are necessarily complex-valued quantities.

In this article, we explain how we can carry out CCM calculations for such 
Hamiltonians that contain  terms in the Hamiltonian after
rotation of local spin axes with complex-valued coefficients. 
We show that the amendments to the existing CCM code 
for spin-lattice models \cite{code}
to be able to treat such cases is non-trivial.  In order to 
illustrate the new technique, we present three separate applications
to models of considerable interest in quantum magnetism.
As a first test of the new methodology we present results  in 
Sec.~\ref{Ising-ferro-chain} for the exactly solvable one-dimensional 
Ising model in a transverse external magnetic field, and an explicit analytical calculation
of the lowest-order implementation of the CCM is 
presented in detail for this model in Appendix \ref{App-A}.  
Secondly, in Sec.~\ref{trianHeis} we then describe results for the spin-half 
Heisenberg model on the triangular lattice at zero temperature in the presence of an 
external magnetic field, in which we make explicit use of the  
3D ``umbrella'' state as our CCM model state,
and an explicit derivation of the Hamiltonian after the rotations of the 
local spin axes for this state is presented in Appendix \ref{App-B}. 
Lastly, in Sec.~\ref{XXZ},  the phase diagram of the 
spin-half {\it XXZ} model on the triangular lattice at zero temperature, also 
in the presence of an external magnetic field (near saturation), is examined.  
Here we again employ the 3D ``umbrella'' state as a possible CCM
model state, and we show how its use now leads to an improved quantitative
description and understanding of this model. We conclude with a brief summary of
our results in Sec.~\ref{summary}.

\section{The Coupled Cluster Method (CCM)}
\label{ccm}

\subsection{Ground-State Formalism}
\label{ccm-A}

As the methodology of the CCM has been discussed extensively elsewhere 
\cite{refc1,refc2,refc3,refc4,refc5,refc6,refc7,refc8,refc9,RFB_1998,ccm1,ccm2,ccm3,ccm4,
ccm5,ccm6,ccm7,ccm7a,ccm8,CCM_shastry,ccm9,ccm10,ccm11,ccm12,ccm13,ccm14,kago1,
ccm15,ccm16,ccm17,ccm18,ccm19,kago2,kago3,bishop2017,bishop2018}, 
only a brief overview of the method is presented here. The 
ground-state Schr\"odinger equations are given by
%%%%%%%%%%%%%%%%%%%%%%%%%%%%%%%%%%
\begin{equation} 
\hat{H} |\Psi\rangle = E_g |\Psi\rangle
\;; \quad
\langle\tilde{\Psi}| \hat{H} = E_g \langle\tilde{\Psi}| 
\;, 
\label{eq1} 
\end{equation} 
%%%%%%%%%%%%%%%%%%%%%%%%%%%%%%%%%%  
in terms of the Hamiltonian $\hat{H}$,
and where formally, for normalisation, we require
\begin{equation}
\langle\tilde{\Psi}| = \frac{(|\Psi\rangle)^{\dag}}{\langle\Psi|\Psi\rangle}\;.
\end{equation}
The bra and ket states for our $N$-spin system (with each of the spins
carrying the spin quantum number $S$) are parametrised {\em independently} in the forms
%%%%%%%%%%%%%%%%%% 
\begin{eqnarray} 
|\Psi\rangle = {\rm e}^{\hat S} |\Phi\rangle \; &;&  
\quad \hat S=\sum_{I \neq 0} {\cal S}_I \hat{C}_I^{+}   \; , 
\label{CCM-ket}\\ 
\langle\tilde{\Psi}| = \langle\Phi| \hat{\tilde{S}} {\rm e}^{-\hat S} \; &;& 
\quad \hat{\tilde{S}} =1 + \sum_{I \neq 0} \tilde{{\cal S}}_I \hat{C}_I^{-} \; ,
\label{CCM-bra} 
\end{eqnarray} 
%%%%%%%%%%%%%%%%%% 
within the normal coupled cluster method, in terms of the multi-configurational
CCM (creation and destruction) correlation operators, $\hat{S}$ and
$\hat{\tilde{S}}$, respectively.  The index $I$ here is a set-index that denotes
a set of lattice sites, $I = \{i_1,i_2,\cdots,i_n;\;n=1,2,\cdots2SN\}$, in which each site
may appear no more than $2S$ times, for reasons we describe below. 
We shall be interested specifically 
in the case of infinite systems, $N \to \infty$.  
Note that $\hat{C}_{0}^{+} \equiv \hat{\mathbbm{1}}$ is defined to be the identity
operator in the many-body Hilbert space, the operators
$\hat{C}_I^{+}$ and $\hat{C}_I^{-} \equiv (\hat{C}_I^{+})^\dag$ are respectively, 
$\forall I \neq 0$,
multispin creation and destruction operators, for clusters of up to $N$ spins, 
which are defined more fully below, 
and ${\cal S}_I $ and $\tilde{{\cal S}}_I$
are the CCM ground-state ket- and bra-state ($c$-number) multispin
correlation coefficients, respectively.
We use model states (denoted $|\Phi\rangle$ for the 
ket state and $\langle\Phi|$ for the bra state) 
as references states for the CCM.  The ket state $|\Phi\rangle$ is required 
to be a fiducial vector (or cyclic vector) with respect to the complete set of 
mutually commuting, multispin creation operators $\{\hat{C}_{I}^{+}\}$. 
Equivalently, the set of states  $\{\hat{C}_{I}^{+}|\Phi\rangle\}$ is a
complete basis for the ket-state Hilbert space.  Furthermore, $|\Phi\rangle$
is also defined to be a generalised vacuum state with respect to
the set of operators $\{\hat{C}_{I}^{+}\}$, in the sense that
%%%%%%%%%%%%%%%%%%%%%%%%%%%%%%%%%%
\begin{equation} 
\langle\Phi|\hat{C}_{I}^{+} = 0 = \hat{C}_{I}^{-}|\Phi\rangle \;; \quad \forall I \neq 0 \;.
\label{vacuum} 
\end{equation} 
%%%%%%%%%%%%%%%%%%%%%%%%%%%%%%%%%%  
We note that, with these conditions fulfilled, the exact ground-state 
ket- and bra-state wave functions, $|\Psi\rangle$ and $\langle\tilde{\Psi}|$,
respectively, now satisfy the normalisation conditions
%%%%%%%%%%%%%%%%%%%%%%%%%%%%%%%%%%
\begin{equation} 
\langle\tilde{\Psi}|\Psi\rangle = \langle\Phi|\Psi\rangle = \langle\Phi|\Phi\rangle \equiv 0.
\label{norms} 
\end{equation} 
%%%%%%%%%%%%%%%%%%%%%%%%%%%%%%%%%%  

We now define the ground-state energy functional, 
$\bar H \equiv  \langle \tilde \Psi | \hat{H} | \Psi \rangle = \langle \Phi | \hat{\tilde S} {\rm e}^{-\hat{S}}\hat{H} {\rm e}^{\hat{S}} | \Phi \rangle$, 
such that the CCM ket- and bra-state equations are given by 
extremising $\bar H$ with
respect to all of the CCM multispin correlation coefficients,
\begin{eqnarray} 
\frac {\partial \bar H} {\partial \tilde{{\cal S}}_I} = 0 \quad  &\Rightarrow& \quad
\langle\Phi|\hat{C}_I^{-} {\rm e}^{-\hat{S}} H {\rm e}^{\hat{S}}|\Phi\rangle = 0 \;,  \;\; 
\forall I \neq 0 \; , \label{ket_state_eqn} \\ 
\frac {\partial \bar H} {\partial {\cal S}_I} = 0  \quad &\Rightarrow&  \quad
\langle\Phi|\hat{\tilde{S}} {\rm e}^{-\hat{S}} [\hat{H},\hat{C}_I^{+}] {\rm e}^{\hat{S}}|\Phi\rangle 
= 0 . \;\; \forall I \neq 0 \; . \label{bra_state_eqn}
\end{eqnarray} 
With these equations satisfied, the CCM ground-state energy is now given by 
\begin{equation}
E_g = \langle \Phi | {\rm e}^{-\hat{S}} \hat{H} {\rm e}^{\hat{S}} | \Phi \rangle \; .
\label{gsenergy}
\end{equation}
Equation (\ref{gsenergy}) is a function of the ket-state correlation coefficients 
\{${\cal S}_I $\} only and 
it involves the similarity transform, ${\rm e}^{-\hat{S}} \hat{H} {\rm e}^{\hat{S}}$,
of $\hat{H}$, which is a key feature of any CCM
calculation. We may evaluate this expression in terms of the well-known 
nested-commutator expansion for the similarity transform of an arbitrary operator $\hat{O}$, 
\begin{equation}
{\rm e}^{-\hat{S}} \hat{O}\, {\rm e}^{\hat{S}} =  \hat{O} + [\hat{O},\hat{S}] + 
\frac 1{2!} [[\hat{O},\hat{S}],\hat{S}] + \cdots \; .
\label{nested-commutator-expansion}
\end{equation}
The Hamiltonian, $\hat{H}$, like any other physical operator
whose CCM ground-state expectation value we wish to calculate,
normally contains only finite sums of 
products of spin operators, and so their nested-commutator expansions
of Eq.\ (\ref{nested-commutator-expansion}) generally terminate 
after a finite number of terms. 

The choice of model state depends 
on the specific details of the model under consideration and so 
this is discussed in detail below. However, we remark that 
a passive rotation of the local spin axes is used in all cases 
such that all spins point in the negative  $z$-direction after rotation 
of the local spin axes. This process allows us to treat all spins equivalently 
and it simplifies the mathematical formulation of the CCM and 
the subsequent derivation of its basic equations, viz., Eqs.\
(\ref{ket_state_eqn}) and (\ref{bra_state_eqn}), very considerably. 
The corresponding multispin creation operators \{$\hat{C}_I^{+}$\} are 
thus defined  with respect to this  CCM model state, such that 
\begin{equation}
 |\Phi\rangle=\bigotimes_{k=1}^N|\downarrow\rangle_{i_k}\; ; \quad 
 \hat{C}_I^+ = \hat{s}_{i_1}^{+} \hat{s}_{i_2}^{+} \cdots \hat{s}_{i_n}^{+}, \; n=1,2, \cdots , 2SN\; ,
\end{equation}
in these rotated local spin-space frames, where $i_k$ denotes an arbitrary lattice site, 
$|\downarrow\rangle_{i_k}$ is the ``downward-pointing'' state
of a spin on site $i_k$ with spin quantum number $S$ (i.e., defined so  that
$\hat{s}^{z}_{i_k}|\downarrow\rangle_{i_k} = -S|\downarrow\rangle_{i_k}$), and 
$\hat{s}_{i_k}^{+} \equiv \hat{s}_{i_k}^{x} + {\rm i} \hat{s}_{i_k}^{y}$ is the usual SU(2) 
spin-raising operator on site $i_k$.

The CCM formalism would be exact if all possible multispin cluster
correlations could be included in the operators $\hat{S}$ and $\hat{\tilde S}$. However,
this is normally impossible to achieve practically. 
In most cases, systematic approximation schemes are used to truncate the 
respective summations in Eqs.\ (\ref{CCM-ket}) and (\ref{CCM-bra}) for these operators,
by restricting the sets of multispin configurations \{$I$\} 
to some manageable subset
within some hierarchical scheme that becomes exact in the limit 
that all configurations are retained.
In the present paper we use two schemes that are denoted as
the SUB$n$--$n$ and LSUB$n$ schemes, respectively. 
The more general SUB$n$--$m$ scheme retains all correlations involving only $n$ or
fewer spin flips (with respect to the respective model state 
$|\Phi\rangle$) that span a range of no more than $m$ contiguous lattice sites. 
By contrast, in the localised LSUB$n$ scheme all multispin correlations over
all distinct locales on the lattice defined by $n$ or fewer contiguous sites are retained.
Each spin flip is defined to require the action of a spin-raising operator 
$\hat{s}_{i_n}^+$ acting just once, and a set of lattice sites is said 
to be contiguous if every site of the set is a nearest neighbour (in some
specified lattice geometry) to at least one other member of the set.
The LSUB$n$ and SUB$n$--$n$ schemes are thus identical only for the limiting case when 
$S = 1/2$. For higher spin quantum numbers $S$, the LSUB$m$
scheme is equivalent to the SUB$n$--$m$ scheme if and only if $n = 2Sm$. 
Spin-cluster configurations $I$ that are equivalent under the
space- and point-group symmetries of the 
crystallographic lattice (as well as of both the Hamiltonian and the model state under consideration) are counted only once 
by explicitly incorporating those symmetries 
into the calculation, and 
these clusters are referred to as ``fundamental clusters''. The number of such 
fundamental clusters used for the ground-state expansions for $|\Psi\rangle$ and
$\langle\tilde{\Psi}|$ at the respective $n$th-order level of (either LSUB$n$ or SUB$n$--$n$)
approximation to Eqs.\ (\ref{CCM-ket}) and (\ref{CCM-bra}) is denoted by $N_f(n)$. 

Although, formally, the CCM correlation operators $\hat{S}$ and
$\hat{\tilde{S}}$ of Eqs.\ (\ref{CCM-ket}) and (\ref{CCM-bra}) must obey the 
condition
\begin{equation}
\langle\Phi|\hat{\tilde{S}}= \frac{\langle\Phi | {\rm e}^{\hat{S}^\dag} {\rm e}^{\hat{S}}}{\langle\Phi | {\rm e}^{\hat{S}^\dag} {\rm e}^{\hat{S}}|\Phi\rangle}\;,
\end{equation}
which is implied by Hermiticity, in practice this may not be exactly fulfilled at finite
levels of (LSUB$n$ or SUB$n$--$n$) approximate implementation, due to 
the independent parametrisations of the two operators.  However, this
minor drawback of the CCM is far outweighed in practice by the two huge advantages
that the method {\em exactly} obeys both the Goldstone linked-cluster theorem
and the very important Hellmann-Feynman theorem at {\em all} levels in the
approximation hierarchies.  The former implies that we can work from the outset
in the required thermodynamic limit of an infinite number of spins, $N \to \infty$, while
the latter implies that the expectation values of all physical parameters are calculated within 
the CCM on the same footing as the energy and in a fully self-consistent manner.

Unlike in many other competing formulations of quantum many-body theory,
the CCM thus never needs any finite-size scaling of the results obtained with it.  Indeed, the 
{\em sole} approximation that is ever made within any application of the CCM is
to extrapolate the results obtained for any physical parameter within the (LSUB$n$ 
or SUB$n$--$n$) approximation hierarchy used to the limit $n \to \infty$ where the
method becomes {\em exact}.  By now, a great deal of experience has been acquired
on how to perform such CCM extrapolations, and we allude here to one such
calculation in Sec.~\ref{Ising-ferro-chain}, and invite the reader to consult the literature cited
above for further details.

\subsection{Computational Aspects for 3D Model States}

The CCM equations (\ref{ket_state_eqn}) and (\ref{bra_state_eqn}) 
may be readily derived and solved analytically at low orders of 
approximation. A full explanation of how this is carried out for the LSUB1 
approximation for the spin-half ferromagnetic Ising chain in a
transverse magnetic field, which we study in Sec.~\ref {Ising-ferro-chain}
is given in Appendix A. Highly 
intensive computational methods  \cite{ccm5,ccm6,ccm7}
are essential at higher orders of LSUB$n$ or SUB$n$--$n$ approximation 
because the number $N_f(n)$
of fundamental clusters (and so therefore also the computational resources 
necessary to store and solve them) scales approximately
exponentially with the order $n$ of the approximation scheme being used. There are 
four distinct steps to perform in carrying out high-order CCM calculations for 
the ground state for ``3D model states,'' 
each of which has a counterpart in the ``standard'' CCM 
code \cite{code} that pertains only to coplanar states.
(As a short-hand only, we shall refer
to {\em any} case that results in the Hamiltonian containing terms with coefficients
that are complex-valued
after rotation of the local spin axes to be a ``3D model state,'' although 
clearly these are some essentially artificial cases, 
e.g., the transverse Ising model presented below, 
where the model state might be coplanar.)   

The first step is to read in ``CCM script files'' that define the basic
problem to be solved. We remark that the derivation of Hamiltonians after rotation
of local spin axes for the 3D model states is non-trivial because we must carry 
out at least two sets of rotations. An example of this process is given for 
the spin-half triangular-lattice Heisenberg model in the presence of an 
external magnetic field in Appendix B. As a consequence the resulting CCM script file 
is much longer than for coplanar model states because we now have
terms in the Hamiltonian with both real and imaginary coefficients. 

The second step involves the 
enumeration of all connected clusters (also called ``lattice animals'') and all disconnected 
clusters that are to be retained at a given level of LSUB$n$ 
or SUB$n$--$n$ approximation for a given lattice and spin quantum 
number $S$ that are distinct under the lattice, model state and Hamiltonian symmetries 
(and perhaps that also satisfy some such conservation rule as $s_T^z=0$,
where $\hat{s}_T^z \equiv \sum_{i=1}^{N} \hat{s}_i^z$,
which would pertain, for example, to all models whose Hamiltonians contain
only spins interacting pairwise via isotropic Heisenberg exchange
interactions). This step is no more difficult for 3D 
model states than for coplanar model states, although clearly this step is 
itself highly non-trivial to perform computationally.

The third step involves deriving and storing the basic CCM ground-state equations. 
In order to find these equations, we first partition the multispin 
cluster configuration pertaining to the set index $I$ for the operator $\hat{C}_I^-$ 
in the ket-state equation $\langle \Phi | \hat{C}_I^- {\rm e}^{-\hat{S}}
 \hat{H} {\rm e}^{\hat{S}} | \Phi \rangle =0$ of Eq.\ (\ref{ket_state_eqn}) into the products 
of ``high-order CCM operators''  \cite{ccm5,ccm6,ccm7}. There are a huge number 
of partitions potentially and each term in a new potential contribution to the 
ket-state equations must be tested for suitability, i.e., all subclusters are checked against 
a list of fundamental clusters after any appropriate space- and point group
symmetries (plus any applicable conservation laws) have been employed.
This is arguably the most difficult step in carrying out any high-order CCM
calculation, and effectively we must run this code twice for the 3D model states:
once for the terms in the Hamiltonian with real coefficients and again for the terms
with imaginary coefficients. 

The fourth step is to solve the ground-state ket and bra equations and to obtain the 
ground-state expectation values. For 3D model states, complex-number algebra 
must be implemented for all subroutines that solve the ket- and bra-state 
equations (solved by ``direct iteration'' for 3D model states), and also in those subroutines
that determine expectation values such as the ground-state energy of Eq.\ (\ref{gsenergy}) 
or other expectation values (i.e., $\bar A = \langle \Phi |\hat{\tilde S} e^{-\hat{S}}
 \hat{A} e^{\hat{S}} | \Phi \rangle$). This is achieved by using options 
in the ${\tt C}$++ compiler. 

Although the process of updating the existing 
high-order CCM code \cite{code} for coplanar states so as to be able
now also to utilise 3D model states 
(resulting in Hamiltonians with terms involving complex-valued coefficients) is 
therefore straightforward in principle, this process is actually considerably less so in practice
because the CCM code \cite{code} itself is extensive and complex. 
In order to validate the new code, we show 
in Sec.~\ref{Ising-ferro-chain} that analytical low-order LSUB1 results 
derived in Appendix A for the transverse Ising model are replicated by the new 3D CCM code.
Similarly, for the same model, for higher orders of LSUB$n$ approximation with $n \leq 12$,
we also show that the new code exactly replicates the
corresponding results obtained using the ``standard'' code \cite{code}.  Both
of these results are excellent tests of the new code.  Furthermore, 
all results for each of the three models 
considered in Secs.\ \ref{Ising-ferro-chain}, \ref{trianHeis} and \ref{XXZ}
are in excellent agreement with the results of other methods (where they exist). 
Finally, we remark again that the creation of the CCM script files
is more complicated for 3D model states than for coplanar model states.

\section{Spin-Half Ising Ferromagnetic Chain in a Transverse External Magnetic Field}
\label{Ising-ferro-chain}

We take as  a first example to demonstrate the feasibility
and the accuracy of the new CCM approach an exactly solvable model, namely
the one-dimensional (1D) spin-$\frac{1}{2}$ Ising ferromagnet in a
transverse magnetic field \cite{pfeuty}.  
The corresponding Hamiltonian is given by
\begin{equation}
\hat{H} = -\sum_{k=1}^{N} \hat{s}_{k}^z \hat{s}_{k+1}^z  - \lambda \sum_{k=1}^{N} \hat{s}_k^x = 
-\sum_{k=1}^{N} \hat{s}_{k}^z \hat{s}_{k+1}^z  - \frac {\lambda}2 \sum_{k=1}^{N} (\hat{s}_k^+ 
+ \hat{s}_k^-) \;,
\label{ising1}
\end{equation}
where the index $k$ runs over all lattice sites on the linear chain (with site $N+1$ 
equivalent to site $1$) and
$\hat{s}_k^{\pm} \equiv  \hat{s}_k^x \pm {\rm i}\hat{s}_k^y$. The strength 
of the applied external {\it transverse} magnetic field is given by $\lambda$.
Clearly, in the case $\lambda = 0$ with no field applied, the spins are
ferromagnetically aligned along the $z$-direction. Similarly, for high enough 
values of $\lambda$ it is clear that the spins will align along the transverse 
($x$-)direction. 
The CCM model state that we choose for this system is one in which all spins point 
in the downwards $z$-direction.  Thus, the model state is expected to be
better for low values of $\lambda$, particularly those below the phase transition
that separates the two regimes where the spins are respectively 
canted to align along some intermediate 
direction between the $z$- and $x$-directions (at low values of $\lambda$) and
fully aligned in the transverse field ($x$-)direction (at high values of $\lambda$).

\begin{figure}
\epsfxsize=15cm
\centerline{\epsffile{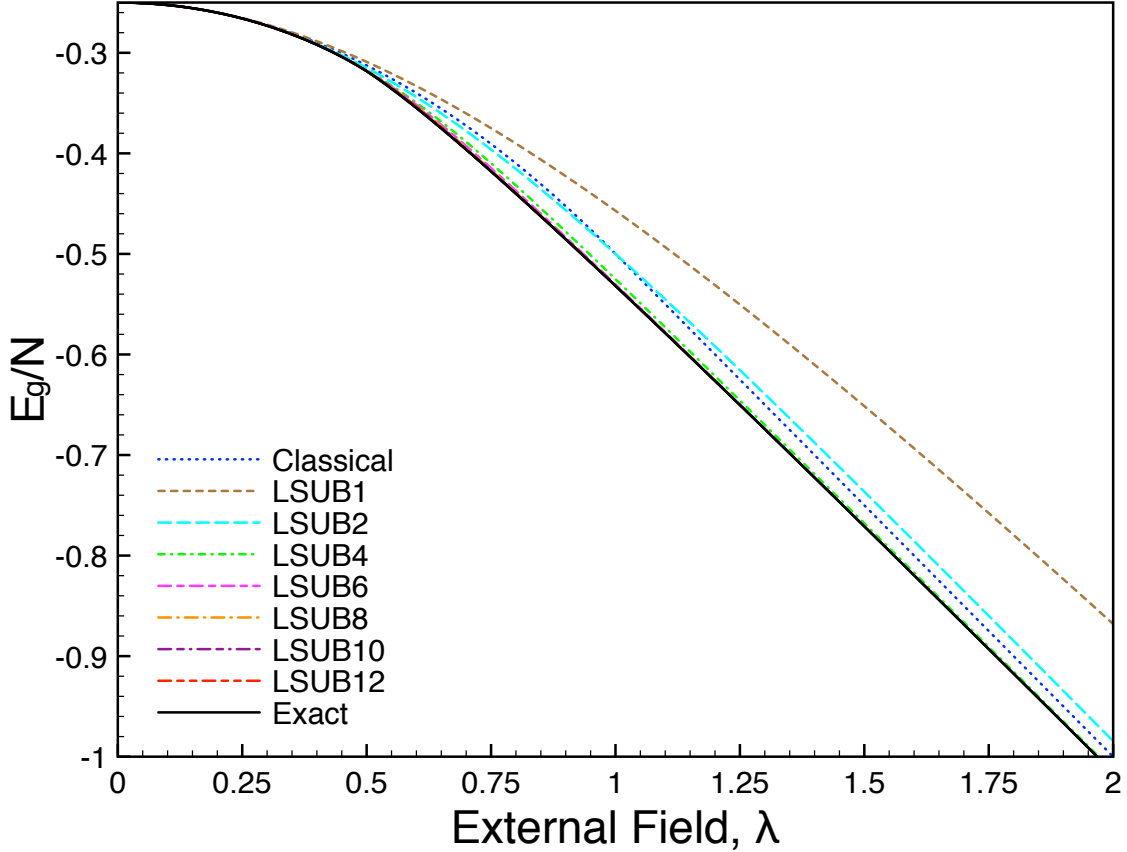}}
\caption{CCM results for the ground-state energy per spin of the Ising model on the linear 
chain as a function of the transverse external magnetic field strength, $\lambda$, 
at various LSUB$n$ levels of approximation.
Also shown are the corresponding classical result of Eq. (\ref{ising-classical-E}) and
the exact result \cite{pfeuty} of Eq.\ (\ref{ising-exact-E}). 
}
\label{ti_energies}
\end{figure}

Classically, the spins are canted at an angle $\alpha$ from the (say, downwards) $z$-direction
in the presence of the transverse magnetic field $\lambda$.  It is trivial to see 
that the classical ground-state energy $E_{g}^{\mathrm{cl}}$ is minimised for 
$\alpha = \sin ^{-1} \lambda$ for  $\lambda \leq 1$.  There is then a classical 
phase transition at $\lambda = \lambda_{c}^{\mathrm{cl}} \equiv 1$, such that 
for $\lambda \geq \lambda_{c}^{\mathrm{cl}}$, the spins are all aligned 
in the direction of the transverse field, with $\alpha = \frac{1}{2}\pi$. We thus have 
that the classical ground-state energy per spin is given by 
\begin{equation}
\frac{E_{g}^{\mathrm{cl}}}{N}=\begin{cases}
                  -\frac{1}{4}(1 + \lambda^{2})\; ; \quad &\lambda \leq 1\\
                  -\frac{1}{2}\lambda\; ; &\lambda \geq 1\; .
                \end{cases}
\label{ising-classical-E}                
\end{equation}
Similarly, the classical values of the magnetisations in the  $z$-direction
(i.e., the Ising direction), $M^z$, and in the transverse $x$-direction 
(i.e., the field direction), $M^{\mathrm{trans.}}$, are trivially found to be given by 
\begin{equation}
M^{z}_{\mathrm{cl}}=\begin{cases}
                  \frac{1}{2}\sqrt{1 - \lambda^2}\; ; \quad &\lambda \leq 1\\
                  0\; ; \quad &\lambda \geq 1\; ,
                \end{cases}
\label{ising-classical-Mz}                
\end{equation}
and
\begin{equation}
M^{\mathrm{trans.}}_{\mathrm{cl}}=\begin{cases}
                  \frac{1}{2}\lambda\; ; \quad &\lambda \leq 1\\
                  \frac{1}{2}\; ; \quad &\lambda \geq 1\; .
                \end{cases}
\label{ising-classical-Mtrans}                
\end{equation}

\begin{figure}
\epsfxsize=15cm
\centerline{\epsffile{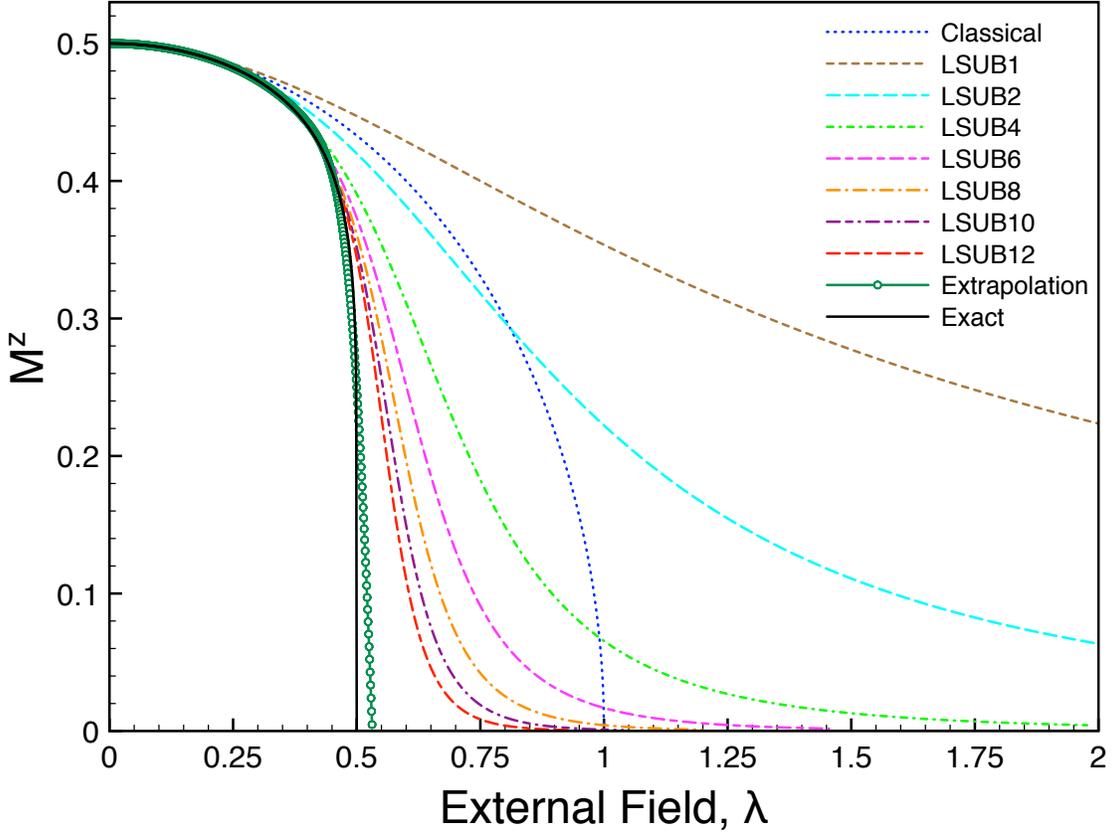}}
\caption{CCM results for the magnetisation $M^z$ of the Ising model on the linear chain  as a function of the transverse external magnetic field strength, $\lambda$,
at various LSUB$n$ levels of approximation, together with the extrapolation based
on all LSUB$n$ results (for both even and odd values
of $n$) with $6 \leq n \leq 12$, as explained in the text.
Also shown are the corresponding classical result of Eq.\ (\ref{ising-classical-Mz})
and the exact result \cite{pfeuty} of Eq.\ (\ref{ising-exact-Mz}). 
}
\label{ti_M}
\end{figure}

\begin{figure}
\epsfxsize=15cm
\centerline{\epsffile{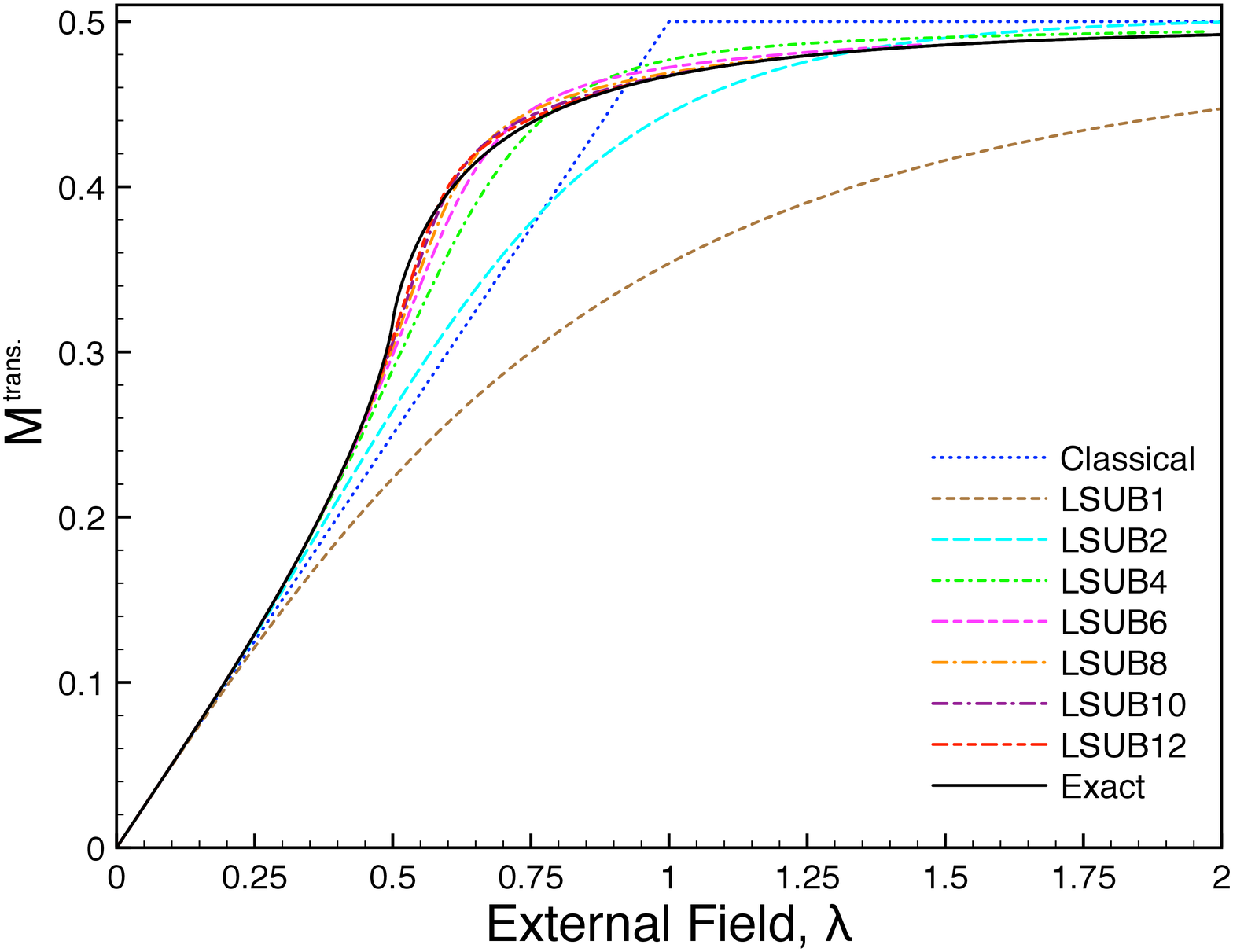}}
\caption{CCM results for the transverse magnetisation, $M^{\mathrm{trans.}}$, of the Ising model on the linear chain as a function of the transverse external 
magnetic field strength, $\lambda$, at various LSUB$n$ levels of approximation. 
Also shown are the corresponding classical result of Eq.\ (\ref{ising-classical-Mtrans}) and
the exact result \cite{pfeuty} of Eq.\ (\ref{ising-exact-Mtrans}).
}
\label{ti_MT}
\end{figure}

The quantum spin-$\frac{1}{2}$ version of the model can be exactly solved \cite{pfeuty}. 
Thus we also have available to us the corresponding exact expressions 
for the classical parameters 
given above in Eqs.\ (\ref{ising-classical-E})--(\ref{ising-classical-Mtrans}), against which 
we can compare our CCM results.  Particular interest attaches to the model due to the 
fact that the classical phase transition at $\lambda = \lambda_{c}^{\mathrm{cl}} \equiv 1$ 
is now shifted to the point $\lambda = \lambda_{c} \equiv \frac{1}{2}$.  The 
exact ground-state energy per spin is given by \cite{pfeuty}
\begin{equation}
\frac{E_{g}}{N} = -\frac{1}{4 \pi} \int_{0}^{\pi} \mathrm{d} k \sqrt{1 + 4 \lambda \cos k +
4{\lambda}^2 }\; ,
\label{ising-exact-E}
\end{equation}
which expression is nonanalytic at the quantum phase transition 
point $\lambda_{c} = \frac{1}{2}$. 
It is simple to check that in the two extremes $\lambda \to 0$ and $\lambda \to \infty$, 
Eq.\ (\ref{ising-exact-E}) reduces respectively to the two limiting values, 
$E_g(\lambda = 0)/N = -\frac{1}{4}$ and
$E_g(\lambda \to \infty)/N \to -\frac{1}{2}\lambda$, exactly as for
the classical case given by Eq.\ (\ref{ising-classical-E}).  This is also just
as expected, since in these two limits the fully aligned ferromagnetic states
are also eigenstates of the quantum Hamiltonian.  Precisely at the quantum
phase transition point, Eq.\ (\ref{ising-exact-E}) yields the value
$E_g(\lambda = \frac{1}{2})/N = -\frac{1}{\pi}$.
The classical result of Eq.~(\ref{ising-classical-E}) is compared with its
exact counterpart of Eq.~(\ref{ising-exact-E}) in Fig.~\ref{ti_energies}.  

The corresponding
exact result for the magnetisation in the (Ising) $z$-direction, $M^z$, is given by \cite{pfeuty}
\begin{equation}
M^z  =\begin{cases}
                  \frac{1}{2}(1 - 4\lambda^2)^{1/8}\; ; \quad &\lambda \leq \frac{1}{2}\\
                  0\; ; \quad &\lambda \geq \frac{1}{2}\; ,
                \end{cases}
\label{ising-exact-Mz}
\end{equation}
which now exhibits the phase transition at $\lambda = \lambda_{c} \equiv \frac{1}{2}$
much more clearly than Eq.\ (\ref{ising-exact-E}) for the ground-state energy.
Once again, the classical and exact results for $M^z$, from 
Eqs. (\ref{ising-classical-Mz}) and (\ref{ising-exact-Mz}) respectively,
are compared in Fig.~\ref{ti_M}.  Finally, the exact result for the transverse
magnetisation (i.e., in the field direction) is given by \cite{pfeuty}
\begin{equation}
M^{\mathrm{trans.}} = \frac{1}{2 \pi} \int_{0}^{\pi} \mathrm{d} k 
\frac{(\cos k + 2 \lambda)}{\sqrt{1 + 4 \lambda \cos k +4{\lambda}^2 }}\; ,
\label{ising-exact-Mtrans}
\end{equation}
which is again nonanalytic at the quantum phase transition point,
 $\lambda_{c} = \frac{1}{2}$, where it takes the value 
 $M^{\mathrm{trans.}}(\lambda = \frac{1}{2})=\frac{1}{\pi}$.  
 It is easy to confirm that when $\lambda$ varies
 from zero to $\infty$, $M^{\mathrm{trans.}}$ from Eq.\ (\ref{ising-exact-Mtrans}) 
 varies smoothly from zero to $\frac{1}{2}$, as shown in Fig.~\ref{ti_MT} 
 where it is also compared to its classical counterpart of Eq.\ (\ref{ising-classical-Mtrans}). 

For present purposes we now wish to illustrate how the CCM can be applied when we 
carry out a unitary transformation of the local spin axes that leads to 
terms in the Hamiltonian with complex-valued coefficients. 
We use the unitary rotation of the local spin 
axes (now for {\it all} sites $k$ on the linear chain) given by 
\begin{equation}
\hat{s}_k^x \rightarrow \hat{s}_k^y \; ; ~~~~ \hat{s}_k^y \rightarrow -\hat{s}_k^x \; ; 
~~~~ \hat{s}_k^z \rightarrow \hat{s}_k^z\;, 
\end{equation}
which simply is equivalent to rotating the transverse field from the $x$- to
the $y$-direction, while leaving the spins aligned in the (negative) $z$-direction.  
That leads to an 
alternative representation of the model given by the Hamiltonian
\begin{equation}
\hat{H} = -\sum_{k=1}^{N} \hat{s}_{k}^z \hat{s}_{k+1}^z  - \lambda \sum_{k=1}^{N} \hat{s}_k^y= 
-\sum_{k=1}^{N} \hat{s}_{k}^z \hat{s}_{k+1}^z  - \frac {{\rm i} \lambda }2 \sum_{k =1}^{N}
(\hat{s}_k^- - \hat{s}_k^+) \; , 
\label{ising2}
\end{equation}
where the term with an imaginary coefficient now appears in the transverse external field
part of the Hamiltonian. However, we remark again that 
the eigenvalue spectrum for this Hamiltonian should not change 
compared to that for the Hamiltonian of Eq.\ (\ref{ising1}).  
Note too that this rotation of the spins in the $xy$-plane does not 
affect the model state for this system, namely, one in which all 
spins point in the downwards $z$-direction.

CCM LSUB1 calculations for both Hamiltonians of Eqs. (\ref{ising1})
and (\ref{ising2}) are carried out explicitly and independently in Appendix A. 
Calculations based on the Hamiltonian of Eq.\ (\ref{ising1}) lead
to ket- and bra-state correlation coefficients that are real numbers only,
whereas those calculations based on the Hamiltonian of Eq.\ (\ref{ising2}) lead
to ket- and bra-state correlation coefficients that are complex (i.e.,
that contain both real and imaginary components). Results
for the ground-state energy per spin, $E_{g}/N$, and the magnetisations, 
$M^z$ in the Ising ($z$)-direction and $M^{\mathrm{trans.}}$ in the
transverse ($x$)-direction, based on the Hamiltonians 
of Eqs. (\ref{ising1}) and (\ref{ising2}) are found to be identical (and so also 
``real-valued'') at the LSUB1 level of approximation, as required.
These analytical results provide a preliminary test of the validity of the CCM 
method for unitary rotations of local spin axes that lead to terms in the Hamiltonian
with both real and imaginary coefficients.

The new code developed here for ``3D model states'' can be applied to the 
Hamiltonian of Eq.\ (\ref{ising2}) to high orders of LSUB$n$ approximation. These 
results can be compared to those from the ``standard'' CCM code \cite{code} 
that works for coplanar states only, which can be applied to the Hamiltonian 
of Eq.\ (\ref{ising1}). Results from these two codes are again found to agree exactly 
with each other at equivalent levels of approximation and specifically also with 
the analytical LSUB1 results presented in Appendix A. The results for the ground-state 
energy are shown in Fig.~\ref{ti_energies} and the results for the 
magnetisations $M^z$ and 
$M^{\rm trans.}$ are shown in Figs. \ref{ti_M} and \ref{ti_MT}, respectively.  
Despite the fact that the ket- and bra-states correlation coefficients are found 
to be complex-valued for all values of $\lambda$  ($>0$), the ground-state
energies and magnetisations are again found to be real  at all approximation 
levels and for all values of $\lambda$ for the Hamiltonian of Eq.\ (\ref{ising2})
using the new code.  

We note that convergence of the LSUB$n$ sequences of approximants becomes worse 
for larger values of $\lambda$ ($\gtrsim 0.5$), exactly as expected, since this region 
is precisely where the model state becomes a poorer
starting point for the CCM calculations, due to the quantum phase transition that
occurs at $\lambda_{c} = \frac{1}{2}$. Nevertheless, it is clear by inspection of 
Figs.~\ref{ti_energies} and \ref{ti_MT}
that results for the ground-state energy and also $M^{\rm trans.}$ compare 
extremely well with the exact results of Ref. \cite{pfeuty} for {\em all}
values of $\lambda$, especially for the 
higher-order LSUB$n$ approximations with $n\gtrsim 6$. 
Although the results for $M^z$ in Fig.~\ref{ti_M} 
also compare well, by inspection,
with the exact results of Ref. \cite{pfeuty}   
in the region where $M^z$ is known to be non-zero from these exact calculations,
(i.e., $\lambda < \frac{1}{2}$.), the agreement is now much poorer outside
this region (i.e., $\lambda > \frac{1}{2}$) for any of the LSUB$n$ approximants shown.  
However, even in this case, the agreement is found to become excellent when
the LSUB$n$ sequence of approximants is extrapolated to the exact limit,
$n \to \infty$, as alluded to in Sec.~\ref{ccm-A}. Thus, a very well-tested
extrapolation scheme for use in such cases where the system undergoes
a quantum phase transition (see, e.g., Ref.\ \cite{bishop2018} and references
cited therein) is
\begin{equation}
M^z(n) = \mu_0 + \mu_1 n^{-1/2} + \mu_2 n^{-3/2}\;,
\label{extrapo}
\end{equation} 
where $M^{z}(n)$ is the $n$th-order CCM approximant (i.e., at
the LSUB$n$ or SUB$n$--$n$ level) to $M^z$.  Thus, in Fig.~\ref{ti_M} we also
show the extrapolation using Eq.\ (\ref{extrapo}) as the fitting formula, 
together with the LSUB$n$ approximants (for both the even values of $n$ shown and the unshown
odd values) with $6 \leq n \leq 12$ as the input data, to determine the extrapolated 
value $\mu_0$, which is plotted.  Clearly, the extrapolation now agrees extremely 
well with the exact result, even in the very sensitive region very close to the
critical value $\lambda_c$, the value of which itself is now also predicted rather accurately.

{\section{Spin-Half Triangular-Lattice Heisenberg Antiferromagnet in an External
Magnetic Field} \label{trianHeis}

We now consider the spin-$\frac{1}{2}$ triangular-lattice
Heisenberg antiferromagnet in a magnetic field.
The Hamiltonian that we will use here is given by 
\begin{equation}
\hat{H} = \sum_{\langle i,j \rangle} \hat{{\bf s}}_i  \cdot  \hat{{\bf s}}_j - 
\lambda \sum_{i=1}^{N} \hat{s}_i^z
\; ,
\label{heisenberg}
\end{equation}
where the index $i$ runs over all $N$ lattice sites on the triangular lattice and 
the sum over the index $\langle i,j \rangle$ indicates a sum over all nearest-neighbour 
pairs, with each pair being counted once and once only. The strength 
of the applied external magnetic field is again given by $\lambda$.
The triangular lattice is itself tripartite, being composed 
of three triangular sublattices,
denoted as $A$, $B$ and $C$, the sites of which we denote 
respectively as $A_n$, $B_n$, and $C_n$. If the original lattice has a
distance $a$ between nearest-neighbour sites, the corresponding distance on 
each of the sublattices $A$, $B$ and $C$ is $\sqrt{3}a$.

It is easy to see that the classical spin-$S$ model 
corresponding to Eq.\ (\ref{heisenberg})
(see, e.g., Ref.~\cite{kawamura})
has an infinitely (and continuously) degenerate 
family of ground states, red with the associated order parameter space
being isomorphic to the 3D rotation group SO(3).  Thus, one may readily rewrite
the classical energy per spin for this model in the form
\begin{equation}
\frac{E^{\mathrm{cl}}}{N} = \frac{1}{4N} \sum_{k=1}^{2N} \left ( \mathbf{S}_{\Delta_k} - \frac{1}{3} \boldsymbol\lambda \right )^2 -\frac{3}{2}S^2 - \frac{1}{18}{\lambda}^2\;,
\label{classical-heisenberg}
\end{equation}
where $\boldsymbol\lambda = \lambda \hat{z}$ and $\hat{z}$ is a unit vector in the 
$z$-direction, and $\mathbf{S}_{\Delta_k} \equiv \mathbf{S}_{A_{\Delta_k}}
+ \mathbf{S}_{B_{\Delta_k}} + \mathbf{S}_{C_{\Delta_k}}$ is 
defined to be the sum of the three spins on
the $k$th elementary triangular plaquette on the lattice with nearest-neighbour vertices
$A_{\Delta_k}$, $B_{\Delta_k}$, and $C_{\Delta_k}$.  Equation (\ref{classical-heisenberg})
shows clearly that the energy is minimised when each of the squared terms in
the sum over elementary triangular plaquettes is either zero (which is possible
for $\lambda \leq 9S$) or minimised [viz., to take the value $(3S-\frac{1}{3}\lambda)^2$ 
for $\lambda > 9S$].  Thus, we find rather simply that the classical ground-state energy 
per spin is given by
\begin{equation}
\frac{E_{g}^{\mathrm{cl}}}{N}=\begin{cases}
                 -\frac{3}{2} S^2 - \frac{1}{18} \lambda^2\; ; \quad &\lambda \leq 9S\\
                 3 S^2 - \lambda S\; ; &\lambda > 9S\; .
               \end{cases}
\label{Heisenberg-triangle-classical-gsE}                
\end{equation}
For $\lambda \leq 9S$, the ground state is clearly infinitely (and continuously)
 degenerate, since any
configuration of spins that satisfies $\mathbf{S}_{\Delta_k} = \frac{1}{3} \boldsymbol\lambda$
on all $2N$ elementary triangular plaquettes will yield the same energy.  Furthermore,
this condition immediately yields that the comparable classical value for the lattice magnetisation
$M$, where $\mathbf{M} \equiv \sum_{i=1}^{N} \mathbf{S}_{i} =M\hat{z}$ 
(i.e., in the direction of the field), in the ground state is given by
\begin{equation}
\frac{M_{\mathrm{cl}}}{S}=\begin{cases}
                  \frac{\lambda}{9S}\; ; \quad &\lambda \leq 9S\\
                  1\; ; \quad &\lambda > 9S\; ,
                \end{cases}
\label{Heisenberg-triangle-classical-M}                
\end{equation}
from which we also see that the magnetisation saturates at the value 
$\lambda = \lambda_s = 9S$ of the magnetic field strength.

In the zero-field case ($\lambda = 0$) the energy-minimising
condition (viz., that $\mathbf{S}_{\Delta_k} = 0$
on all $2N$ elementary triangular plaquettes) simply becomes the condition
for the usual $120^\circ$ three-sublattice N\'{e}el state. Associated with
this state there is clearly a trivial degeneracy due to the rotational
invariance of any Hamiltonian composed only of isotropic Heisenberg interactions,
which is reflected in the ground-state order parameter space being
isomorphic to the group SO(3).  In the case of a finite external field
($\lambda \neq 0$) the symmetry of the Hamiltonian of Eq.\ (\ref{heisenberg})
is clearly reduced from SO(3) to SO(2)$\times  \mathrm{Z}_3$,
corresponding to the rotational symmetry around the axis of the magnetic field
and the discrete symmetry associated with the choice of the three sublattices
$A$, $B$ and $C$.  Despite this reduction in symmetry of the finite-field
($\lambda \neq 0$) Hamiltonian of Eq.\ (\ref{heisenberg}) from that
of its zero-field ($\lambda = 0$) counterpart, the ground state of the former clearly
shares the same [i.e., SO(3)] degree of continuous degeneracy as that of the latter,
due to the condition that $\mathbf{S}_{\Delta_k} = \frac{1}{3} \boldsymbol\lambda$
on all $2N$ elementary triangular plaquettes. Thus, on each plaquette, each of the 
three spins has two orientational degrees of freedom, and the above condition 
simply reduces the overall degrees of freedom from six to three.
The trivial degeneracy of the $\lambda = 0$ case is now, however, quite
non-trivial in the $\lambda \neq 0$ case, since the local $120^\circ$ 
triangular-plaquette structures can become quite deformed by the
application of the external field, even into non-coplanar configurations,
as we now discuss.

From our discussion above, in principle, depending on the magnetic field strength, any of the five ground-state spin configurations sketched in  Fig.~\ref{model_states} may appear.
\begin{figure}[t!]
\epsfxsize=16cm
\centerline{\epsffile{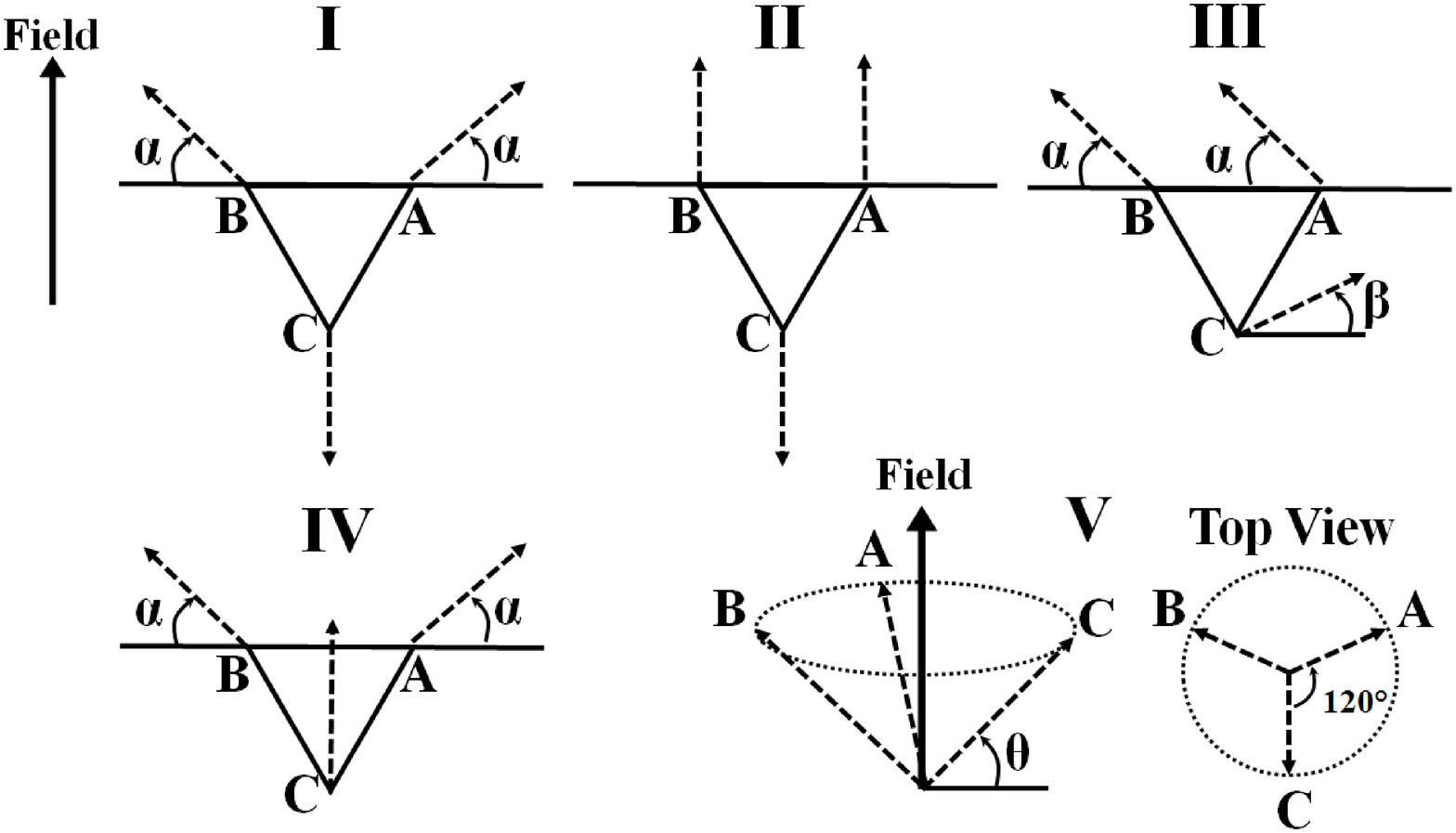}}
\caption{Some examples of possible (degenerate) classical ground states 
(and hence also possible CCM model states) of
the spin-half triangular-lattice Heisenberg antiferromagnet in 
an external magnetic field: states I to IV are coplanar, 
whereas state 
V is the non-coplanar (3D) ``umbrella'' state 
with spins at an angle $\theta$ to the plane perpendicular 
to the external field.}
\label{model_states}
\end{figure}
While the states I, II, III and IV are coplanar states, the
``umbrella'' state V is a 3D non-coplanar state.  
Although on the classical level both coplanar and 
non-coplanar sates are energetically degenerate, 
%\cite{kawamura,chub,zhito}, 
as we have noted above for $\lambda \leq \lambda_s$,
thermal fluctuations tend to
favour the coplanar configurations \cite{kawamura,chub,zhito,zhito_kago}. 

By minimising the energy, it is easy to show that the classical spin-$S$ model 
described by Eq.\ (\ref{heisenberg}) has a (coplanar)
ground state of type I in Fig.~\ref{model_states} for $\lambda < 3S$, with a
canting angle $\alpha$ given by
\begin{equation}
\sin \alpha = \frac{1}{2} + \frac{\lambda}{6S}\;;\quad \lambda \leq 3S.
\end{equation}
At zero field ($\lambda = 0$) state I simply becomes the usual $120^\circ$ 
three-sublattice N\'{e}el state, while
precisely at the value $\lambda =3S$ the state I becomes the collinear state II
shown in Fig.~\ref{model_states}, and
as $\lambda$ is increased further the ground state now smoothly transforms into
state III shown in Fig.~\ref{model_states}.  The canting angles $\alpha$ and $\beta$
are found to be given by
\begin{equation}
\sin \alpha = \frac{(\lambda^2 + 27S^2)}{12 \lambda S}\;;\quad 3S \leq \lambda \leq 9S,
\end{equation}
and
\begin{equation}
\sin \beta = \frac{(\lambda^2 - 27S^2)}{6 \lambda S}\;;\quad 3S \leq \lambda \leq 9S,
\end{equation}
which may readily be shown to satisfy the condition, $2 \cos \alpha = \cos \beta$,
which ensures that state III does not acquire any lattice magnetisation transverse
to the applied field.
When the field strength takes the value $\lambda = 3S$ the angles are 
$\alpha = \frac{1}{2} \pi$ and $\beta = -\frac{1}{2} \pi$, which is again just
equivalent to state II.
As $\lambda$ is then increased, up to the saturation value $\lambda=\lambda_s  =9S$,
the angle $\alpha$ first decreases to its minimum value, $\alpha=\frac{1}{3} \pi$,
at $\lambda = 3\sqrt 3 S$, after which it again increases smoothly back to the value
$\alpha = \frac{1}{2} \pi$ at $\lambda = 9S$.  At the same time, as 
$\lambda$ is increased beyond the value $3S$, the angle $\beta$
increases from $-\frac{1}{2} \pi$ to $\frac{1}{2} \pi$ at $\lambda = 9S$, taking
the value $\beta = 0$ in between, precisely at the point $\lambda = 3\sqrt 3 S$
where $\alpha$ becomes a minimum.  For all values $\lambda > \lambda_s = 9S$ the
ground state is the fully saturated ferromagnetic state (viz., state III with
$\alpha = \frac{1}{2} \pi = \beta$).  One may readily show that the classical ground-state
energy of Eq.\ (\ref{heisenberg}), for both states I and III at the respective values of
their minimising canting angles and for the fully saturated state, is just
that given previously in Eq.\ (\ref{Heisenberg-triangle-classical-gsE}).
Furthermore, the corresponding classical value for the lattice magnetisation (i.e., in the direction 
of the field) in the ground states I and III is given by our previous result of Eq.\ (\ref{Heisenberg-triangle-classical-M}).

Although the state IV shown in Fig.~\ref{model_states} is not utilised as a CCM
model state in any further application in this Section to the spin-$\frac{1}{2}$ 
case of the Hamiltonian of Eq.\ (\ref{heisenberg}), it will be 
considered later in Sec.~\ref{XXZ}.  Hence, for completeness, we note that state IV
has a minimum energy for the classical spin-$S$ case of the present Hamiltonian 
for a value of the canting angle $\alpha$ shown in Fig.~\ref{model_states} given by
\begin{equation}
\sin \alpha = - \frac{1}{2} + \frac{\lambda}{6S}\;;\quad \lambda \leq 9S.
\label{Heisenberg-triangle-classical-state-IV-angle} 
\end{equation}
Hence, unlike the situation for state I, which undergoes a 
smooth transformation to 
state III at a value, $\lambda = \frac{1}{3} \lambda_s$, of the external 
field strength, (which then itself smoothly varies as $\lambda$
is further increased up to the value $\lambda_s$, at which point
it becomes the fully saturated ferromagnetic state), state IV simply 
varies smoothly from the $120^\circ$ 
three-sublattice N\'{e}el state at zero field, $\lambda = 0$, to the fully saturated
ferromagnetic state at $\lambda = \lambda_s$.

For the classical case, as we have already noted, a non-coplanar 
state of the form of state V of Fig.~\ref{model_states} is degenerate in energy 
with states I and III above
in their respective regimes.  Thus, one readily finds that state V has a minimum
energy for the classical spin-$S$ Hamiltonian of Eq.\ (\ref{heisenberg}) for a value of
the out-of-plane angle $\theta$ given by
\begin{equation}
\sin \theta = \begin{cases} 
    \frac{\lambda}{9S}\;;\quad & \lambda \leq 9S\\
    1\; ; \quad &\lambda > 9S\; .
                    \end{cases}
\label{Heisenberg-triangle-classical-state-V-angle}                    
\end{equation}
Thus, with that value of $\theta$, state V also yields a value for the energy identical to 
that of Eq.\ (\ref{Heisenberg-triangle-classical-gsE}).  Clearly, the lattice magnetisation
is then also given by Eq.\ (\ref{Heisenberg-triangle-classical-M}). 

For the  quantum spin-$\frac{1}{2}$ case, no exact solution is available,
but many investigations
\cite{LhuiMi,BLLP,hon1999,CGHP,squareTriangleED,HSR04,yoshikawa,nishi,chub,alicea,ccm12,ono,fortune,sakai2011,seabra2011,hotta2013,ccm17,zhito2014,coletta}
have demonstrated that the {\it order from disorder} mechanism
\cite{villain,shender} selects coplanar spin configurations, and, in
particular, a wide magnetisation plateau 
at one-third of the saturation value (i.e., at $\lambda = \frac{1}{3}\lambda_s$)  is present
\cite{LhuiMi,BLLP,hon1999,CGHP,squareTriangleED,HSR04,yoshikawa,nishi,
chub,alicea,ccm12,ono,fortune,sakai2011,seabra2011,hotta2013,ccm17,zhito2014,coletta}.
This is precisely the value of the field strength for which the collinear
state II is degenerate with other ground-state spin configurations.  Since it is 
well known that quantum fluctuations tend to favour collinear over non-collinear
spin configurations, it is no real surprise that in the extreme quantum limiting case
$S=\frac{1}{2}$ the classical transition point at $\lambda = \frac{1}{3}\lambda_s$
should broaden into a plateau.
A previous investigation using the CCM \cite{ccm12} with the coplanar states I,
II and III as the model states showed that for the spin-$\frac 1 2$ 
model the  plateau state occurs for  $1.37 \lesssim \lambda \lesssim 2.15$. 
Here our aim is to compare new CCM results generated with the 3D ``umbrella'' state
V as the model state to those obtained previously for the coplanar states \cite{ccm12}.

According to the CCM scheme briefly outlined in
Sec.~\ref{ccm}
we have to perform a passive rotation of the local spin
axes of the spins such that all spins appear to point downwards for 
all five model states I, II, III, IV and V in Fig.~\ref{model_states}. 
For the coplanar model states I, II, and III that procedure has been explained and discussed 
in detail in Ref.~\cite{ccm12}.  A similar rotation is also necessary for the
coplanar  model state IV, that does not play a role in this section, but 
which will be  used in Sec.~\ref{XXZ}.   
For the non-coplanar model state V, that comprises spins 
that make an angle $\theta$ to the plane perpendicular to the external field, 
the derivation of the Hamiltonian after rotation 
of the local spin axes is given in Appendix B, from which we 
note that the final result is given by
\begin{equation}
\begin{split}
\hat{H} &=
\sum_{\langle i_{B,C,A} \rightarrow j_{C,A,B}\rangle} \biggl \{  
\left(\sin^2 \theta - \frac{1}{2} \cos^2 \theta \right) \hat{s}_{i_{B,C,A}}^z \hat{s}_{j_{C,A,B}}^z \\ 
&  \qquad \qquad \qquad 
+ \frac{1}{4} \left( \frac{1}{2} \sin^2 \theta - \cos^2 \theta -  \frac{1}{2} \right) \bigl (\hat{s}_{i_{B,C,A}}^+ \hat{s}_{j_{C,A,B}}^+ + \hat{s}_{i_{B,C,A}}^- \hat{s}_{j_{C,A,B}}^- \bigr )
\\ & \qquad \qquad \qquad 
+ \frac{1}{4} \left(\cos^2 \theta -  \frac{1}{2} \sin^2 \theta - \frac{1}{2} \right) \bigl (\hat{s}_{i_{B,C,A}}^+ \hat{s}_{j_{C,A,B}}^- + \hat{s}_{i_{B,C,A}}^- \hat{s}_{j_{C,A,B}}^+ \bigr ) \\ &  \qquad \qquad \qquad 
+ \frac{\sqrt{3}}{4}  \cos \theta \, \Bigl [ \hat{s}_{i_{B,C,A}}^z \bigl (\hat{s}_{j_{C,A,B}}^+ + \hat{s}_{j_{C,A,B}}^- \bigr ) - \bigl ( \hat{s}_{i_{B,C,A}}^+ + \hat{s}_{i_{B,C,A}}^- \bigr ) \hat{s}_{j_{C,A,B}}^z \Bigr ]
\biggr \} \\ 
 &  \quad \quad
 + \lambda \sum_{i=1}^{N}  \sin \theta \, \hat{s}_{i}^z  \\ 
 & \quad \quad
 + {\rm i}   \sum_{\langle i_{B,C,A} \rightarrow j_{C,A,B}\rangle} \biggl \{  
 \frac{\sqrt{3}}{4}  \sin \theta \, \bigl ( \hat{s}_{i_{B,C,A}}^- \hat{s}_{j_{C,A,B}}^+ - \hat{s}_{i_{B,C,A}}^+ \hat{s}_{j_{C,A,B}}^-\bigr ) \\ 
 &  \qquad \qquad \qquad  
+ \frac{3}{4} \sin \theta  \cos\theta \, \Bigl [\hat{s}_{i_{B,C,A}}^z \bigl ( \hat{s}_{j_{C,A,B}}^+ - \hat{s}_{j_{C,A,B}}^- \bigr ) + \bigl ( \hat{s}_{i_{B,C,A}}^+ - \hat{s}_{i_{B,C,A}}^- \bigr ) \hat{s}_{j_{C,A,B}}^z\Bigr ]
 \biggr \}  \\
& \quad \quad
+ {\rm i}\,   \frac{\lambda}{2} \sum_{i=1}^{N} \cos \theta \, \bigl ( \hat{s}_{i}^+ - \hat{s}_{i}^- \bigr ) \;,
    \label{rotH4-text} 
\end{split}    
\end{equation}
where the sums over $\langle i_{ B,C,A} \rightarrow j_{C,A,B}\rangle$ represent 
a shorthand notation to include the  three sorts of ``directed'' nearest-neighbour 
bonds on each basic triangular plaquette
of side $a$ on the triangular lattice, which join sites $i_B$ and $j_C$ going from the 
$B$-sublattice to the $C$-sublattice, sites $i_C$ and $j_A$ going from the $C$-sublattice to the 
$A$-sublattice, and sites $i_A$ and $j_B$ going from the $A$-sublattice to the $B$-sublattice 
(in those directions only and not reversed). We see that this 
Hamiltonian now contains terms with both real and imaginary coefficients. 

Clearly, when using any classical configuration of spins as a CCM
model state, such as those shown in Fig.~\ref{model_states}, there is no 
reason to expect that the quantum spin-$S$ version of the model, with
a finite value of the spin quantum number $S$,
will take the same values of the angle parameters that characterise it as the
classical version (i.e., in the $S \to \infty$ limit), even in the case that the quantum
ground state (at least partially) preserves the classical ordering inherent 
in the model state.  For this reason a first step in using any such model state
in a CCM calculation is to optimise the angle parameters that characterise 
the spin configuration.  To do so we simply choose those parameters that 
minimise the ground-state energy at each (either LSUB$n$ or SUB$n$--$n$)
level of approximation that we undertake, performing a separate such
optimisation at each level.

\begin{figure}[t!]
\epsfxsize=15cm
\centerline{\epsffile{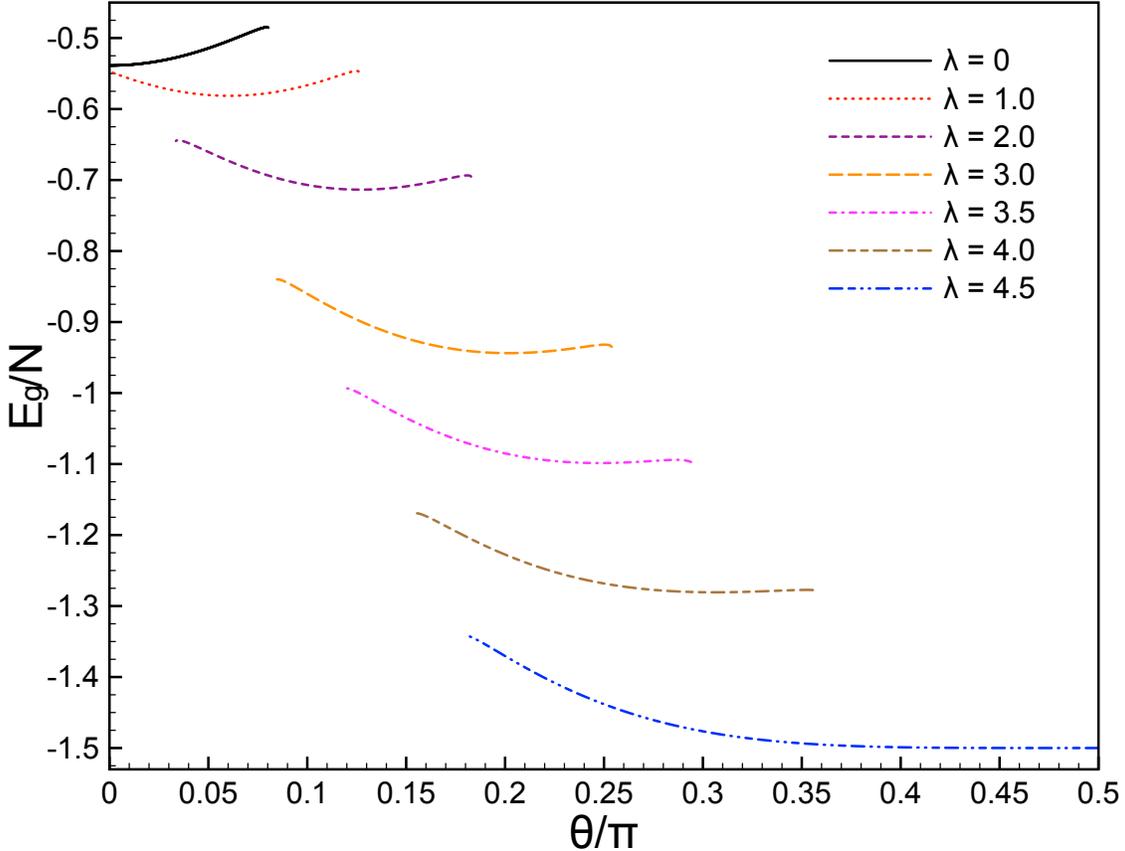}}
\caption{
CCM results for the ground-state energy per site, $E_g/N$, 
of the spin-half triangular-lattice Heisenberg antiferromagnet, calculated at the LSUB5
level of approximation,
plotted as a function of the out-of-plane angle $\theta$ (in units of
$\pi$) for the 3D non-coplanar state
V, shown for various values of the 
external magnetic field strength, $\lambda$ in the range between zero
and the saturation value $\lambda_s = \frac{9}{2}$.
}
\label{angle}
\end{figure}

\begin{figure}
\epsfxsize=15cm
\centerline{\epsffile{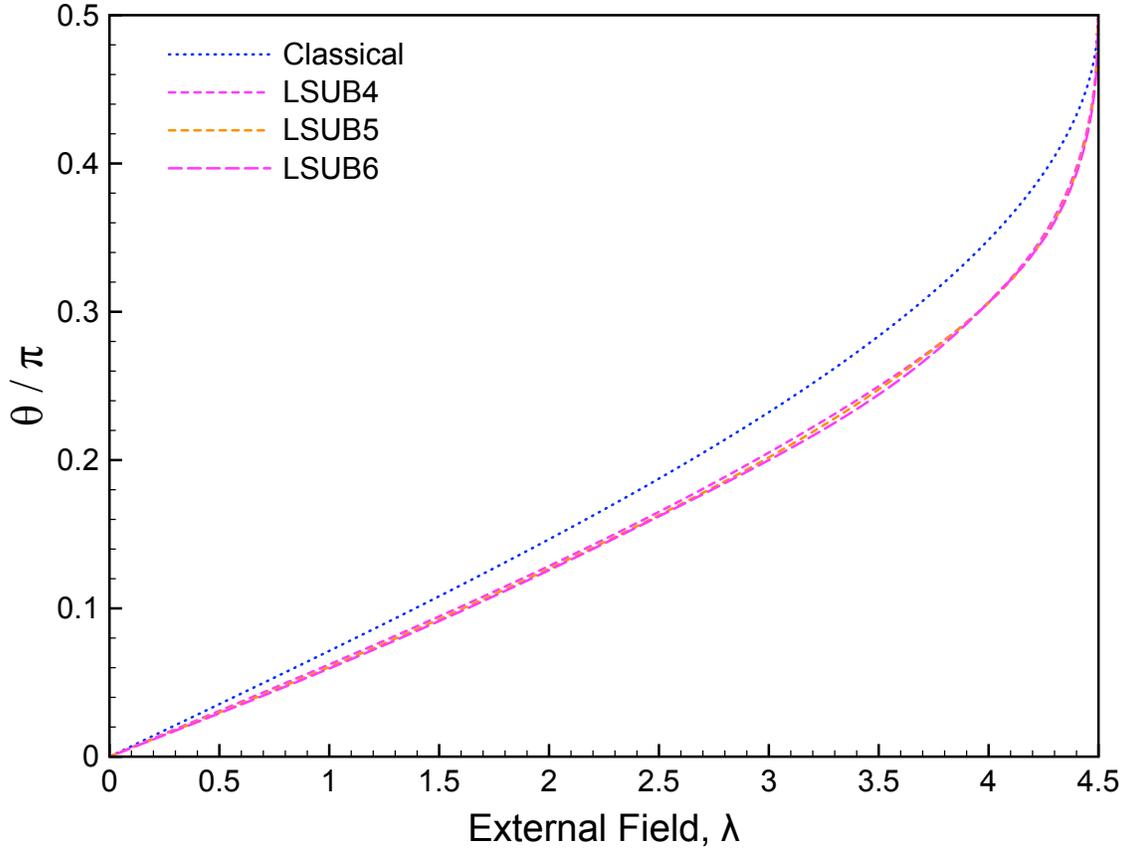}}
\caption{
CCM results for the out-of-plane angle $\theta$ (in units of $\pi$) that minimises the ground-state 
energy of the spin-half triangular-lattice Heisenberg antiferromagnet in
an external magnetic field of strength  
$\lambda$, plotted as a function of $\lambda$ for the 3D non-coplanar state V,
at various LSUB$n$ levels of approximation. For
comparison purposes we also show the corresponding classical result from 
Eq.\ (\ref{Heisenberg-triangle-classical-state-V-angle}) with $S=\frac{1}{2}$.
}
\label{angle2}
\end{figure}

Typical such CCM results for the ground-state energy of the spin-$\frac{1}{2}$ 
Hamiltonian described by Eq.\ (\ref{heisenberg}) from using the non-coplanar
state V as the model state are shown in Fig.~\ref{angle} 
as a function of the 
out-of-plane angle $\theta$ and for various values of the external field strength 
in the range $0 \leq \lambda \leq \lambda_s$, for the particular case of
the LSUB5 level of approximation.
The ground-state energy is found to be a real number 
for all values of $\lambda$ and $\theta$. We note in 
particular that all purely imaginary contributions to the 
energy sum identically to zero.
Figure \ref{angle} demonstrates the general result that the ground-state energy
has a well-defined minimum with respect to the angle  
for all values of $\lambda$, at each LSUB$n$ level of approximation. 
The angle that minimises the 
ground-state energy is plotted 
as a function of $\lambda$ in Fig.~\ref{angle2} for various levels of LSUB$n$ approximation. 
As required, this angle is zero (i.e., the model state is coplanar) when the 
external field is zero ($\lambda = 0$). Also as required, all spins 
point in the direction of the field (i.e., shown by 
$\theta/\pi=\frac{1}{2}$) 
at the saturation field, $\lambda_s= \frac{9}{2}$. The angle that 
minimises the energy varies continuously as a function of 
$\lambda$ for model state V, excepting the limiting point at 
``saturation'', $\lambda_s$.  Furthermore, we see from 
Fig.~\ref{angle2} that the non-coplanar energy-minimising configuration of spins
converges very rapidly as the LSUB$n$ approximation index $n$ is increased, for all 
values of $\lambda$.

\begin{figure}
\epsfxsize=15cm
\centerline{\epsffile{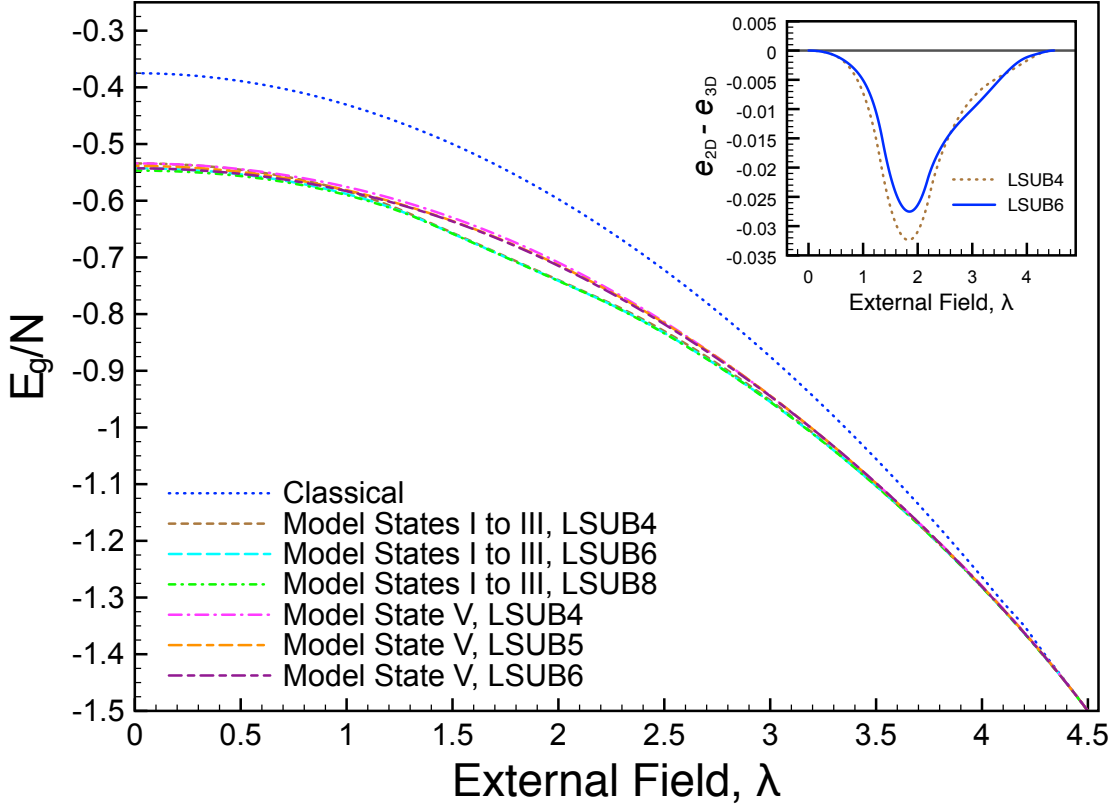}}
\caption{
Main: CCM results for the ground-state energy per site, $E_g/N$, of the 
spin-half triangular-lattice Heisenberg antiferromagnet in
an external magnetic field of strength  
$\lambda$, plotted as a function of $\lambda$, using 
both coplanar states I-III (results from Ref. \cite{ccm12}) and the 3D non-coplanar 
state V as CCM model states, at various LSUB$n$ levels of approximation. For
comparison purposes we also show the corresponding classical result from 
Eq.\ (\ref{Heisenberg-triangle-classical-gsE}) with $S=\frac{1}{2}$.
Inset: Energy difference, $\delta e = e_\mathrm{2D}-e_\mathrm{3D}$ (where
$e \equiv E_{g}/N$), 
between the 2D coplanar states and the 3D
non-coplanar state. 
}
\label{triangle_energies}
\end{figure}

Ground-state energies for model state V 
are shown in Fig.~\ref{triangle_energies}, in which
results for the coplanar model states I to III from 
Ref. \cite{ccm12} are also shown for comparison. Again, LSUB$n$ results for the energy 
converge rapidly with increasing levels of the truncation index $n$
for all values of $\lambda$. We see very clearly that ground-state energies 
for the coplanar model states lie lower than those of the 
3D ``umbrella'' state (model state V) for all values $\lambda$,
which is in agreement with the results of other methods 
\cite{kawamura,chub,zhito}.
Note that the energy difference between the coplanar and the 
non-coplanar states is particularly large in the plateau region around $\lambda
= 1.5$, as can be seen clearly from the inset in Fig.~\ref{triangle_energies}.

Naturally, the (physical) lattice magnetisation is defined in terms of the spin directions 
{\it before} all rotations of the local spin axes have been carried out. 
Thus, the lattice magnetisation is given in terms of the ``unrotated''
coordinates as:
\begin{equation}
M = \frac 1N \sum_{i=1}^{N} \bigl \langle \tilde \Psi  \bigl | \hat{s}_i^z \bigr | \Psi \bigr \rangle \; .
\label{lmag}
\end{equation}
After the rotations of the local spin axes for model state V, 
which led to the expression of Eq.\ (\ref{rotH4-text}),
have been completed, this expression is given by
\begin{equation}
M = - \frac 1N \sum_{i=1}^{N} \Big \langle \tilde \Psi \Big |  \Bigl (\sin \theta\,  \hat{s}_{i}^z  +  \frac{\rm i }{2}  \cos \theta \,[\hat{s}_{i}^+ - \hat{s}_{i}^-]  \Bigr ) \Big | \Psi \Big \rangle \; .
\label{lmag2}
\end{equation}
Previous initial results for model state V \cite{ccm13} 
used computational differentiation and the Hellmann-Feynman theorem 
to evaluate this lattice magnetisation of Eq.\ (\ref{lmag}). Here we 
evaluate it directly by finding both the ket- and bra-state correlation 
coefficients, which are now complex-valued, and then evaluating 
the expectation value explicitly, although we note that this method 
provides identical results (within the precision allowed by numerical 
differentiation) to that of the former technique, as required. 

\begin{figure}
\epsfxsize=15cm
\centerline{\epsffile{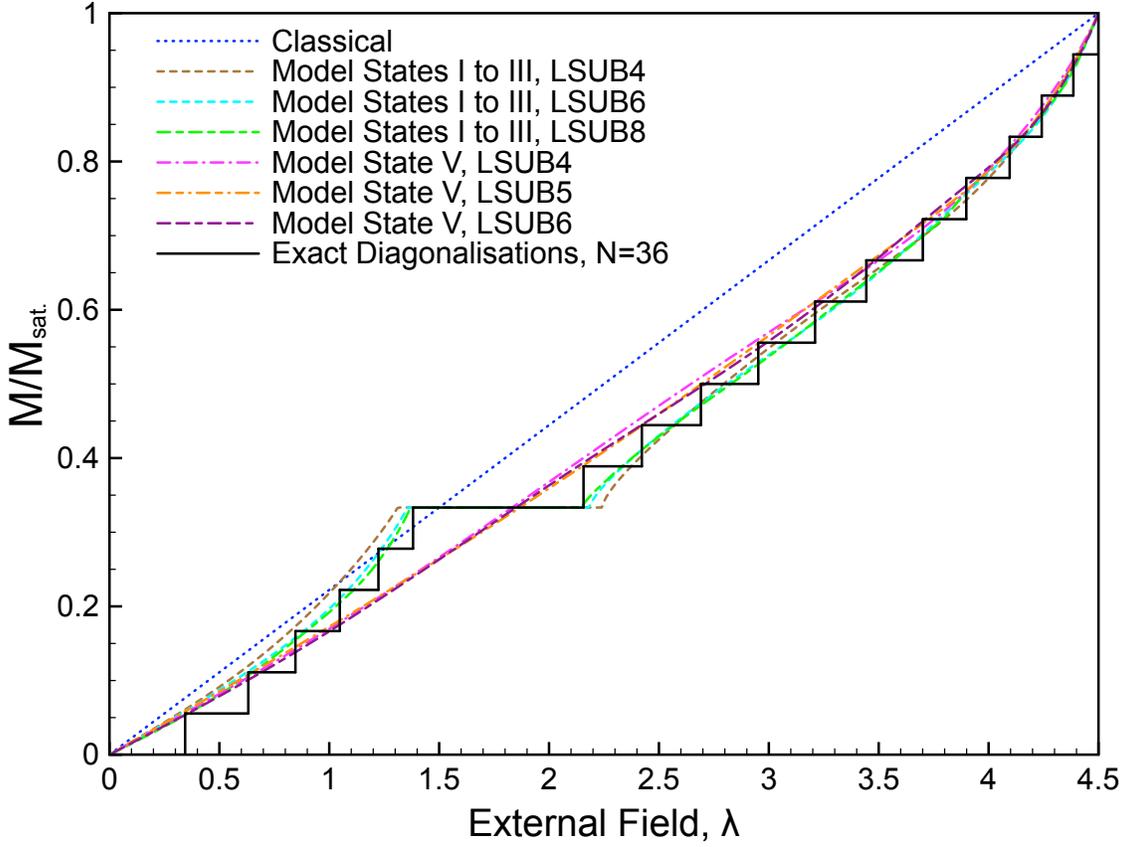}}
\caption{CCM results at various levels of LSUB$n$ approximation for the 
ratio, $M/M_{\rm sat.}$, of the lattice magnetisation to its saturated value, of the 
spin-half triangular-lattice Heisenberg antiferromagnet in
an external magnetic field of strength  
$\lambda$, plotted as a function of $\lambda$, using 
both coplanar states I-III (results from Ref. \cite{ccm12}) and the 3D 
non-coplanar model state V as CCM model states.  For
comparison purposes we also show the corresponding classical result from 
Eq.\ (\ref{Heisenberg-triangle-classical-M}), as well as the result from
an exact diagonalisation of the Hamiltonian on a (36-site) finite-sized lattice.
}
\label{lat_mag}
\end{figure}

Once again, the results for the lattice magnetisation shown in Fig.~\ref{lat_mag} 
for model state V
are found to be real numbers for all values of $\lambda$, with 
all imaginary contributions summing identically to zero.  Furthermore,
LSUB$n$ results are again found to converge with increasing 
approximation level $n$ for all values of $\lambda$. It is evident that CCM results 
for the 3D ``umbrella'' model state V do not indicate the presence of the 
well-known magnetisation plateau that occurs in this system. By contrast, 
results for the coplanar states agree well with those results of
exact diagonalisations, including the well-known plateau regime at 
$M/M_{\rm sat.}=1/3$.  The high-order LSUB8 approximation for the coplanar states, for
example, yields \cite{ccm12} that this regime extends over the region 
$1.37 \lesssim \lambda \lesssim 2.15$, and it is clear too from
Fig.~\ref{lat_mag} that the borders of the plateau region also
converge rapidly as the order $n$ of the LSUB$n$ approximation is increased. 
Such CCM results \cite{ccm12} for the plateau
 for the coplanar states that we have now
shown explicitly lie lower in energy than the 3D ``umbrella'' state, 
are in excellent agreement with experimental
results for the magnetic compound Ba$_3$CoSb$_2$O$_9$ 
(a spin-half triangular-lattice antiferromagnet) and 
exact diagonalisations \cite{shirata}.
Note also that previous CCM results for the coplanar model states
also indicate that a similar plateau occurs over 
the range $2.82  \lesssim  \lambda  \lesssim  3.70$ for the for the spin-one 
triangular-lattice antiferromagnet, and this theoretical result has subsequently 
been established experimentally for the compound Ba$_3$NiSb$_2$O$_9$ 
(a spin-one triangular-lattice antiferromagnet) \cite{ccm17}.

\section{Spin-$S$ Triangular-Lattice {\it XXZ} Antiferromagnet in an External Magnetic Field}
\label{XXZ}

In recent investigations 
 \cite{yamamoto2014,starykh2014,sellmann2015,mamorini2016,yamamoto2017}  of the anisotropic triangular-lattice {\it XXZ} model
\begin{equation}
\hat{H} = \sum_{\langle i,j \rangle} \left ({\hat s}^x_i {\hat s}^x_j+{\hat s}^y_i {\hat s}^y_j 
+ \Delta {\hat s}^z_i{\hat s}^z_j \right )- \lambda \sum_{i=1}^{N} \hat{s}_i^z \; ,
\label{xxz}
\end{equation}
where the indices have the same meaning as in Eq. (\ref{heisenberg}),
it has been shown that for an easy-plane anisotropy (i.e., $\Delta < 1$) the 3D
``umbrella'' state V discussed in Sec.~\ref{trianHeis} can  become energetically
favoured over the coplanar states, so as to form the true ground state under
certain conditions that we now elaborate.

The corresponding phase diagram in the $\Delta$--$\lambda$ plane is rich,
(see, e.g., Ref.~\cite{yamamoto2014}). Moreover, 
the phase boundary between the coplanar and non-coplanar ground states  strongly depends on the spin quantum number $S$. We note that in 
the classical limit $S \to \infty$ for $\Delta < 1$ the  non-coplanar ``umbrella'' state is always energetically favoured over the planar states to form the 
ground state, as we elaborate further below,
whereas for the extreme quantum case $S=\frac 12$ there is a wide region of 
values of the anisotropy  parameter $\Delta$ and the field 
strength $\lambda$ where coplanar states are favoured.
We note further that since
the energy differences between competing ground states can
be very small, accurate and self-consistent calculations are
hence required to be able to distinguish between them reliably.

By comparison with the derivation of Eq.\ (\ref{classical-heisenberg}) for
the Hamiltonian of Eq.\ (\ref{heisenberg}) in Sec.~\ref{trianHeis}, it is clear
that we may write the classical energy per spin for the current 
anisotropic triangular-lattice {\it XXZ} model in the form
\begin{equation}
\frac{E^{\mathrm{cl}}}{N} = \frac{1}{4N} \sum_{k=1}^{2N} \left ( \mathbf{S}_{\Delta_k} - \frac{1}{3} \boldsymbol\lambda \right )^2 + (\Delta -1) \sum_{\langle i,j \rangle}S_i^z S_j^z
-\frac{3}{2}S^2 - \frac{1}{18}{\lambda}^2\;.
\label{classical-XXZ}
\end{equation}
It is evident that the second sum in Eq.\ (\ref{classical-XXZ}) can now potentially favour
non-coplanar states in the case of easy-plane anisotropy (i.e., when $\Delta < 1$).
One readily finds, by making use of Eq.\ (\ref{classical-XXZ}), 
that state V of the form shown in Fig.~\ref{model_states} has a minimum
energy for the classical spin-$S$ Hamiltonian of Eq.\ (\ref{xxz}) for a value of
the out-of-plane angle $\theta$ given by
\begin{equation}
\sin \theta = \begin{cases} 
    \frac{\lambda}{\lambda_s}\;;\quad & \lambda \leq \lambda_s\\
    1\; ; \quad &\lambda > \lambda_s\; ,
                    \end{cases}
\label{XXZ-triangle-classical-state-V-angle}                    
\end{equation}
where $\lambda_s \equiv 3(1 + 2\Delta)S$ is the value of the field strength that 
reaches saturation (i.e., the fully aligned ferromagnetic state) for this state V. With
this value of $\theta$ one may readily show that state V yields a value for the
classical ground-state energy per spin given by
\begin{equation}
\frac{E_{g}^{\mathrm{cl;V}}}{N}=\begin{cases}
                 -\frac{3}{2} S^2 - \frac{\lambda^2}{6(2\Delta + 1)} \; ; \quad &\lambda \leq \lambda_s \\
                 3 \Delta S^2 - \lambda S\; ; &\lambda > \lambda_s \; ,
               \end{cases}
\label{XXZ-triangle-classical-gsE-stateV}                
\end{equation}
which also replicates Eq.\ (\ref{Heisenberg-triangle-classical-gsE}) at isotropy 
(i.e., when $\Delta = 1$).  Furthermore, one may readily show that state V, with
the energy-minimising value of the out-of-plane angle $\theta$ of 
Eq.\ (\ref{XXZ-triangle-classical-state-V-angle}), yields a classical value for the
lattice magnetisation given by
\begin{equation}
\frac{M_{\mathrm{cl;V}}}{S}=\begin{cases}
                \frac{\lambda}{\lambda_s}\; ; \quad &\lambda \leq \lambda_s\\
                  1\; ; \quad &\lambda > \lambda_s\; .
                \end{cases}
\label{XXZ-triangle-classical-M-stateV}                
\end{equation}

We may compare the above results for the non-coplanar state V with
those of the coplanar states.  For example, one may readily show, again by making 
use of Eq.\ (\ref{classical-XXZ}), 
that state IV of the form shown in Fig.~\ref{model_states} has a minimum
energy for the classical spin-$S$ Hamiltonian of Eq.\ (\ref{xxz}) for a value of
the canting angle $\alpha$ given by 
\begin{equation}
\sin \alpha = \begin{cases} 
    \frac{-\Delta + \lambda/(3S)}{(1 + \Delta)}\;;\quad & \lambda \leq \lambda_s\\
    1\; ; \quad &\lambda > \lambda_s\; ,
                    \end{cases}
\label{XXZ-triangle-classical-state-IV-angle}                    
\end{equation}
where $\lambda_s = 3(1 + 2\Delta)S$ as before.  Once again, this result is in accord 
with our previous result of Eq.\ (\ref{Heisenberg-triangle-classical-state-IV-angle}) 
at the isotropic point, $\Delta = 1$.  With
this value of $\alpha$ one may show that state IV yields a value for the
classical ground-state energy per spin given by
\begin{equation}
\frac{E_{g}^{\mathrm{cl;IV}}}{N}=\begin{cases}
                 \frac{1}{(\Delta + 1)} [-(\Delta^2 + \Delta + 1)S^2 + \frac{1}{3}(\Delta - 1)\lambda S - \frac{1}{9} \lambda^2 ]  \; ; \quad &\lambda \leq \lambda_s \\
                 3 \Delta S^2 - \lambda S\; ; &\lambda > \lambda_s \; ,
               \end{cases}
\label{XXZ-triangle-classical-gsE-stateIV}                
\end{equation}
which also replicates Eq.\ (\ref{Heisenberg-triangle-classical-gsE}) at isotropy 
(i.e., when $\Delta = 1$).  One may also show that state IV, with
the energy-minimising value of the canting angle $\alpha$ of 
Eq.\ (\ref{XXZ-triangle-classical-state-IV-angle}) yields a classical value for the
lattice magnetisation given by
\begin{equation}
\frac{M_{\mathrm{cl;IV}}}{S}=\begin{cases}
                \frac{(9S + 4\lambda - \lambda_s)}{3(3S+\lambda_s)}\; ; \quad &\lambda \leq \lambda_s\\
                  1\; ; \quad &\lambda > \lambda_s\; ,
                \end{cases}
\label{XXZ-triangle-classical-M-stateIV}                
\end{equation}
which may be compared with the corresponding result of 
Eq.\ (\ref{XXZ-triangle-classical-M-stateV}) for state V.

Finally, at the classical level, one may readily show from Eqs.\
(\ref{XXZ-triangle-classical-gsE-stateV}) and (\ref{XXZ-triangle-classical-gsE-stateIV})
that the difference between the minimum energy per spin for the ``umbrella'' state 
V and that for the coplanar state IV is given explicitly by
\begin{equation}
\frac{E_{g}^{\mathrm{cl;IV}}}{N} - \frac{E_{g}^{\mathrm{cl;V}}}{N} =
\frac{(1 - \Delta)}{18(\Delta + 1)(2\Delta + 1)}(\lambda - \lambda_s)^2\;;
\quad \lambda \leq \lambda_s.
\label{XXZ-triangle-energy-diff}
\end{equation}
It is evident from Eq.\ (\ref{XXZ-triangle-energy-diff}) that the ``umbrella' state V
always lies lower in energy than the coplanar state IV for all values of the
anisotropy parameter $\Delta < 1$, as we have already asserted, in the classical limit
$S \to \infty$.  Equation (\ref{XXZ-triangle-energy-diff}) shows clearly, however,
that the energy difference between the states decreases both as $\lambda$ approaches 
the saturation value $\lambda_s$ and as $\Delta$ approaches unity 
(i.e., near the Heisenberg isotropic limit).  One expects on rather general grounds
that quantum fluctuations will favour phases with coplanar over non-coplanar configurations
of spins.  One also expects that the effects of quantum fluctuations will
increase monotonically as the spin quantum number $S$ is decreased smoothly 
from the large-$S$ classical limit.  Hence, for small positive values of $(\lambda_s - \lambda)$
it is, {\it a priori}, likely that the value $\Delta_{1}(S)$ of the anisotropy parameter at 
which a possible quantum phase transition occurs between the ``umbrella'' state forming
the ground state (for $\Delta < \Delta_1$) to a coplanar state forming the ground
state (for $\Delta > \Delta_1$) would decrease smoothly as $S$ decreases from the 
classical value $\Delta_{1}(\infty)=1$ at $S \to \infty$ towards
a lower value $\Delta(\frac{1}{2})$ at the extreme quantum limit, $S = \frac{1}{2}$.
  
Thus, to demonstrate the capability of the new 3D CCM code to detect such
anisotropy-driven quantum transitions between 
coplanar and non-coplanar states we now 
consider the same model described by the Hamiltonian of
Eq.\ (\ref{xxz}) for values of the field strength near to the saturation value,
$\lambda_{s}=3(1+2\Delta)S$, but for various finite values of the 
spin quantum number, viz., $S=\frac{1}{2},1,\cdots,5$.
The local spin rotations are identical  to those discussed in Sec.~\ref{trianHeis} and 
Appendix B for model states I, II, III, and V, and are hence not repeated here. 
Model state IV uses the same rotations as for model state I for the $A$ and $B$ 
sublattices, although the ``up'' spins on the $C$ sublattice for this model state 
obviously also require an additional rotation of $180^{\circ}$.

\begin{figure}
\epsfxsize=14.25cm
\centerline{\epsffile{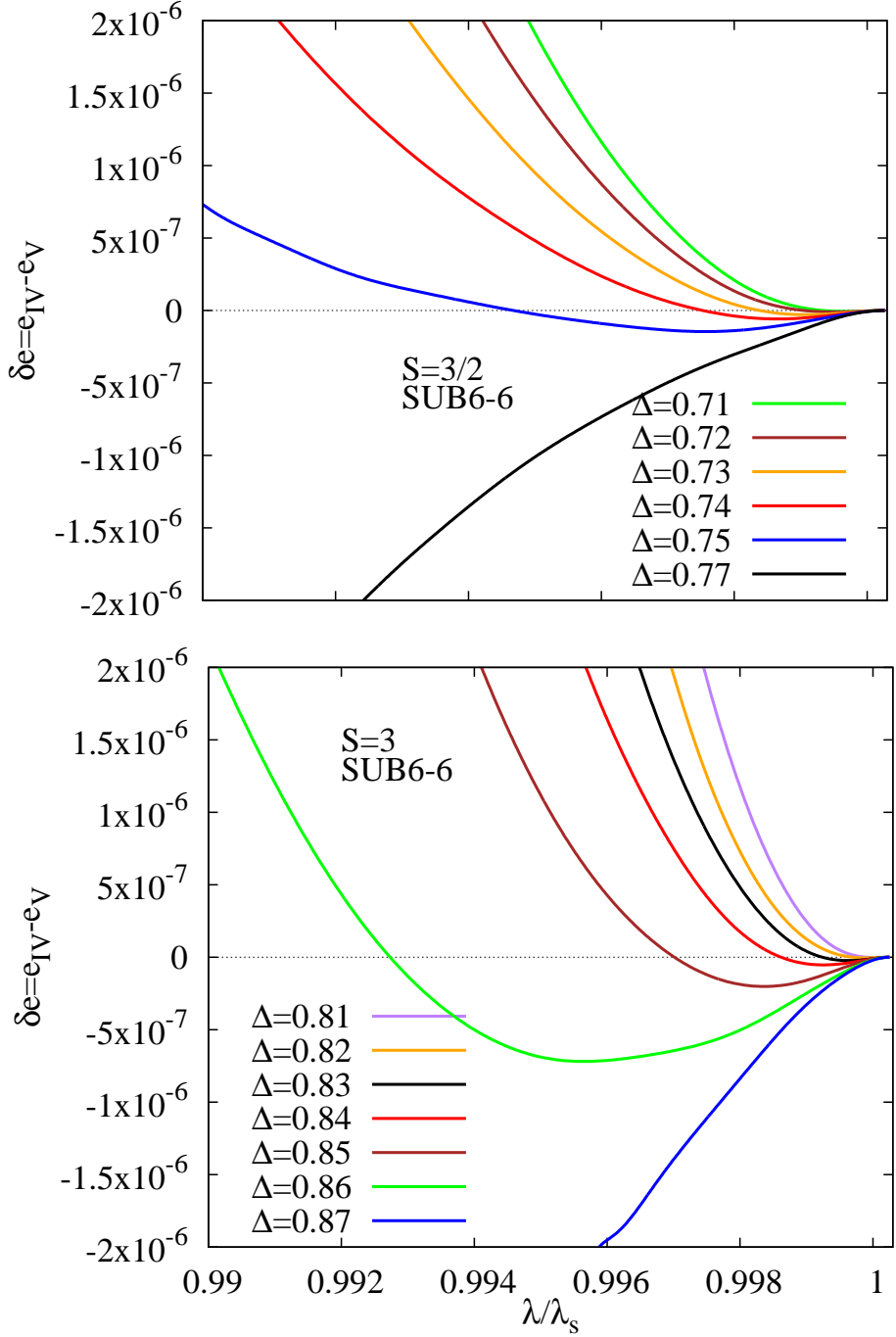}}
%\centerline{\epsffile{e_diff_embed.eps}}
\caption{CCM SUB6-6 results for the difference of energies per spin, 
$\delta e=e_{\rm{IV}}-e_{\rm{V}}$, between the
two competing states IV and V, of a spin-$S$ anisotropic triangular-lattice $XXZ$ 
antiferromagnet in an external magnetic field of
strength $\lambda$, plotted as a function of $\lambda$ (in units of $\lambda_s$), 
for values of $\lambda$ just below
the saturation field strength, $\lambda_s=3S(1+2\Delta)$, for various values 
of the anisotropy parameter, $\Delta$, and 
for two values of the spin quantum number, $S=3/2$ (top panel) 
and $S=3$ (bottom panel).
}
\label{e_diff}
\end{figure}

\begin{figure}
\epsfxsize=15cm
\centerline{\epsffile{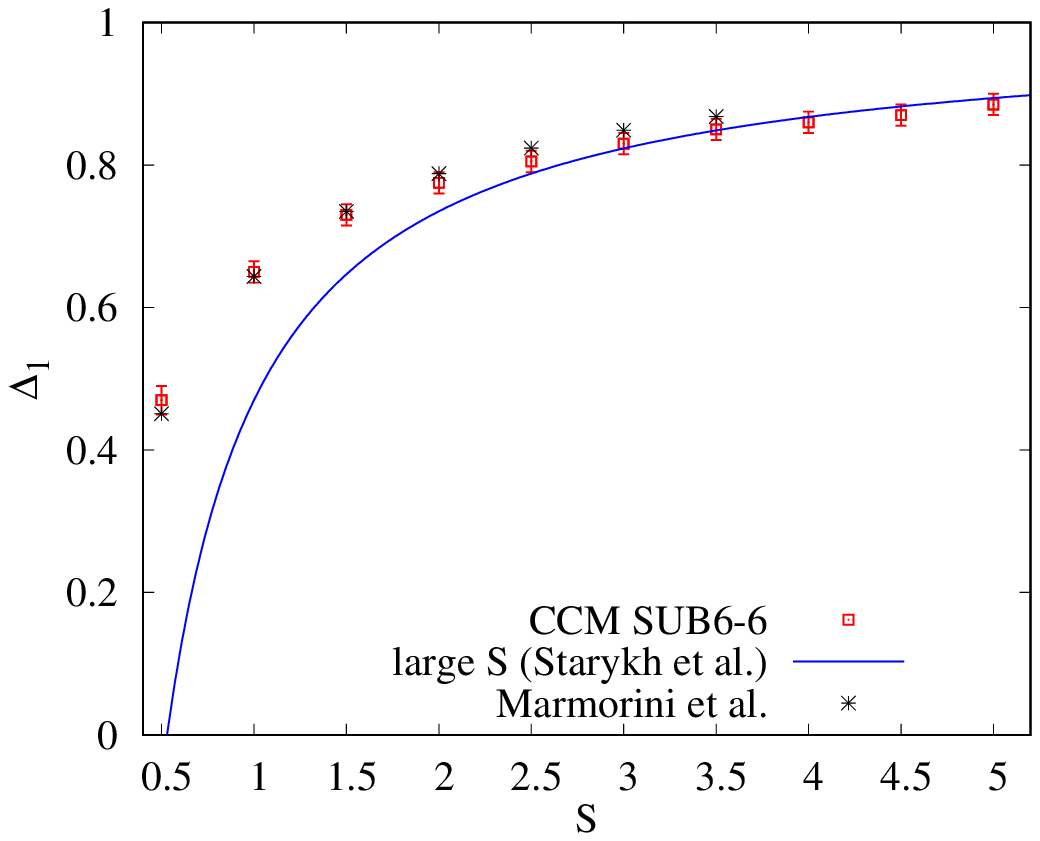}}
\caption{CCM SUB6-6 results with error bars as shown 
for the critical value $\Delta_{1}$ 
of the anisotropy parameter $\Delta$, which denotes the point at 
which the stable ground state changes
from being state V (for $\Delta < \Delta_1$) 
to state IV (for $\Delta > \Delta_1$), of a spin-$S$ 
anisotropic triangular-lattice $XXZ$ 
antiferromagnet in an external magnetic field of
strength infinitesimally below the saturation field strength, 
$\lambda_s=3S(1+2\Delta)$, 
versus the spin quantum number $S$.
The (blue) solid line corresponds to the large-$S$ expansion result, $\Delta_{1}=1-0.53/S$, of
Ref. \cite{starykh2014}.
The (black) stars correspond to the dilute Bose gas expansion results of
Ref. \cite{mamorini2016}.
}
\label{Delta_crit}
\end{figure}

Other investigations \cite{yamamoto2014,starykh2014,sellmann2015,mamorini2016,yamamoto2017} have 
shown that for strong magnetic fields with a strength $\lambda$ infinitesimally below the saturation field strength $\lambda_{s}$ the states III, IV or V 
shown in Fig.~\ref{model_states} appear as the ground state,
depending on the value of the anisotropy parameter $\Delta$.
In the $XY$ limit, $\Delta=0$, the 3D ``umbrella'' state V is always the ground
state, for all values of the spin quantum number $S$. 
Increasing $\Delta$ then leads first to a transition to the coplanar state IV
at a critical value $\Delta_{1}(S)$ and then to a second transition at $\Delta_{2}(S)$      
to  the coplanar state III. Both transition points depend on the value of $S$.
To find the transition points with our CCM approach is straightforward but
computationally quite intensive, since for each spin quantum number $S$ the 
energies of the competing ground states have to be computed for a fine
net of $\Delta$ values, for each of which the corresponding quantum  pitch angles 
that minimise the respective energies at the particular SUB$n$--$n$ level
of approximation being utilised must be determined
iteratively.  Moreover, the size of the set of coupled nonlinear 
CCM ket-state equations increases rapidly with the truncation index $n$, such that 
for the highest LSUB6--6 level of approximation considered here the number 
of such equations is $N_{f}(6)=80339$.
Therefore, we focus particular attention here on the transition between 
the 3D ``umbrella'' state V and  the coplanar state IV, which we have
already discussed above in the classical limit, $S \to \infty$.

In Fig.~\ref{e_diff} we show two examples of the  energy difference
$\delta e=e_{\rm{IV}}-e_{\rm{V}}$ between the
two competing states IV and V, where $e \equiv E/N$ 
represents the energy per spin in each case.
The change of the sign of $\delta e$, as $\Delta$ is varied across the critical
value $\Delta_1$, is obvious.
Note that a change of the sign of $\delta e$ with decreasing $\lambda$ at
fixed $\Delta$ (see, e.g., the green line for $\Delta=0.86$ in the lower panel of
Fig.~\ref{e_diff}), does not actually indicate a reentrance of the ``umbrella''
state V, since for decreasing $\lambda$ the coplanar state III, which we have
not considered here, actually
becomes the ground state (see, e.g., Refs. 
\cite{yamamoto2014,starykh2014,mamorini2016,yamamoto2017}).
Based on curves such as those shown in Fig.~\ref{e_diff} we 
derive the critical points  $\Delta_{1}(S)$   
with an accuracy of $\pm 0.015$, as shown in Fig.~\ref{Delta_crit}.
We compare our CCM data with 
the large-$S$ approach of Starykh et al.\
\cite{starykh2014} which yields
$\Delta_{1}(S)=1-0.53/S$, as well as with numerical data 
obtained by the dilute Bose gas expansion \cite{mamorini2016}.
As can be clearly seen, there is good agreement of our results 
with both those of Ref.~\cite{mamorini2016}
and those of Ref.~\cite{starykh2014}, except where the latter results from 
the large-$S$ expansion
naturally fail for the smaller values of $S$.  Finally, we note too
that our results also agree extremely well with those 
from a recent study that used the numerical cluster mean-field plus 
scaling method \cite{yamamoto2017} to investigate the cases with 
$S \leq \frac{3}{2}$.
To conclude, we have shown that the new CCM code for 3D non-coplanar ground
states provides accurate data that, in particular, allow us to examine with confidence
the quantum selection
of competing ground states of the frustrated triangular-lattice {\it XXZ}
antiferromagnet in an external magnetic field.

\section{Summary}
\label{summary}
Coplanar model states for applications of the CCM 
to problems in quantum magnetism are those states 
in which all spins lie in a plane, whereas 3D model states are 
non-coplanar states in which the spins do not lie in any plane. The first step in 
applying the CCM to lattice quantum spin systems is always to rotate the local 
spin axes (i.e., on each lattice site) so that 
all spins in our model state appear mathematically to points in the downwards (i.e., negative)
$z$-direction.  We have shown explicitly here that this process leads inevitably to
terms in the rotated Hamiltonian with complex-valued coefficients for 3D model states,
by contrast with the case for coplanar model states, where all of the respective
terms carry (or can be made to carry) real-valued coefficients.  Since these rotations represent
unitary transformations the rotated Hamiltonians in each case are still Hermitian.  
Nevertheless, for the case of 3D model states, even though the expectation values 
of all physical operators are hence guaranteed to be real-valued quantities,
the intervening CCM bra- and ket-state 
multispin cluster coefficients from which they will be calculated will be complex-valued at 
all levels  of approximation. 

In this paper we have demonstrated that the existing high-order CCM 
code \cite{code} can be extended appropriately to be able to be
applied to such rotated Hamiltonians with complex-valued coefficients that
arise from the utilisation of 3D model states. An 
explicit derivation of such a Hamiltonian after all rotations of spin axes for a 
3D model state was given for the triangular-lattice Heisenberg antiferromagnet 
in an external magnetic field. Although in such cases the
CCM ket- and bra-state equations for the multispin cluster
correlation coefficients are now complex-valued quantities, we have demonstrated
explicitly for several models of interest that all expectation 
values are real numbers.  
This was also shown explicitly in analytical LSUB1 calculations for the
one-dimensional spin-$\frac{1}{2}$ Ising ferromagnet in a transverse external magnetic field. 

Due to the length and complexity 
of the CCM code, its extension from coplanar to 3D model 
states is a non-trivial task. Furthermore, the task of defining the problem to 
be solved by the code in CCM ``script files'' becomes more difficult. Hence, this 
is an important advance for practical applications of the CCM, and one 
that greatly extends its range of applicability. 

Finally, excellent correspondence with the results of other methods was 
found for all of the cases considered here, namely, (a) the 1D spin-$\frac{1}{2}$ Ising 
ferromagnetic chain in a transverse external magnetic field; (b) the 2D spin-$\frac{1}{2}$
Heisenberg antiferromagnet on a triangular lattice
in the presence of an external magnetic  field; and (c) the 2D spin-$S$
triangular-lattice {\it XXZ} antiferromagnet 
in the presence of an external magnetic field, for the cases $\frac{1}{2} \leq S \leq5 $.
For the first case of the transverse Ising model, which simply provides a 
testbed for which we can artificially
introduce terms in a rotated Hamiltonian pertaining to it that contain
an imaginary coefficient, we showed that CCM results 
agree well with exact results.  With the extended CCM methodology thereby
validated we turned to two frustrated 2D models defined on a triangular lattice, 
for both of which a real 3D non-coplanar configuration of spins is physically relevant.

Our CCM results for the spin-$\frac{1}{2}$ triangular-lattice
Heisenberg antiferromagnet show that coplanar ordering is favoured over 
non-coplanar ordering for all values of the applied external magnetic field, 
which agrees with the results of other approximate methods 
\cite{kawamura,chub,zhito}. The boundary between a 3D ground state (state V) and a
coplanar ground state (state IV) was also obtained for the {\it XXZ} model  
on the triangular lattice 
for values of the external magnetic field  near to saturation and for spin quantum 
number $S \le 5$. The CCM calculations in this case were computationally intensive for this 
frustrated model, especially for high spin quantum numbers. However, 
we note that only a very few other approximate methods can deal effectively 
and accurately with the combination of strong frustration and higher 
spin quantum number. The differences in ground-state energies between the two states is 
also very small in the limit of field saturation.  Hence, an accurate 
delineation of the phase boundary is an extremely delicate 
task. Despite this inherent difficulty excellent correspondence was seen with the results of other 
approximate methods. These results thereby constitute a useful advance in the 
understanding of this model, as well as providing an excellent quantitative test 
of high-order CCM using 3D model states.

%\pagebreak 
\appendix

\section{LSUB1 Calculation for the 1D Spin-$\frac{1}{2}$ Ising Ferromagnet in a Transverse Magnetic Field}
\label{App-A}

We first carry out a CCM LSUB1 calculation for the spin-$\frac{1}{2}$ 
1D transverse Ising model of 
Eq.\ (\ref{ising1}). We recall that the CCM model state contains 
spins that point in the downward $z$-direction. We note also that this is the 
exact (albeit trivial) ground state when $\lambda=0$ and that all CCM 
correlation coefficients should therefore tend to zero in this limit, 
$\lambda \rightarrow 0$. The LSUB1 ket-state operator $\hat S$ is thus given by
\begin{equation}
\hat{S} = a \sum_{i=1}^{N} \hat{s}_i^+ \label{sub1}\; .
\end{equation}
By making use of the usual SU(2) spin commutation relations,
$[\hat{s}_l^+ , \hat{s}_{l'}^-] = 2\hat{s}_l^z \delta_{l,l'}$ 
and $[\hat{s}_l^z , \hat{s}_{l'}^{\pm}] = \pm \hat{s}_l^{\pm} \delta_{l,l'}$, 
and by also using the nested commutator expansion of 
Eq.\ (\ref{nested-commutator-expansion}),
it is readily proven that the CCM similarity transforms of the operators 
$\hat{s}_l^+$, $\hat{s}_l^z$, and $\hat{s}_l^-$ are given by 
\begin{eqnarray}
{\rm e}^{-\hat{S}}\hat{s}_l^+{\rm e}^{\hat{S}} &=& \hat{s}_l^+ \; ;\nonumber \\
{\rm e}^{-\hat{S}}\hat{s}_l^z{\rm e}^{\hat{S}} &=& \hat{s}_l^z +a\hat{s}_l^+ \; ; \\
{\rm e}^{-\hat{S}}\hat{s}_l^-{\rm e}^{\hat{S}} &=& \hat{s}_l^- - 2a \hat{s}_l^z - a^2 \hat{s}_l^+ \nonumber \; ,
\label{sub1_expansions} 
\end{eqnarray}
at the LSUB1 level of approximation. 
Thus we see that the corresponding CCM LSUB1 similarity transform
of the Hamiltonian of Eq.\ (\ref{ising1}) is given by
\begin{equation}
{\rm e}^{-\hat{S}}\hat{H}{\rm e}^{\hat{S}}= -\sum_{i=1}^{N} \left (\hat{s}_{i}^z \hat{s}_{i+1}^z  +  a \hat{s}_{i}^z \hat{s}_{i+1}^+ +  
a \hat{s}_{i}^+ \hat{s}_{i+1}^z + a^2 \hat{s}_{i}^+ \hat{s}_{i+1}^+   \right ) 
-\frac {\lambda}2 \sum_{i=1}^{N} \left (\hat{s}_i^+ + \hat{s}_i^- - 2a \hat{s}_i^z - a^2 \hat{s}_i^+ \right ) \; .
\label{ising_tilde_1}
\end{equation}
By making use of the relation, $\hat{s}^z|\Phi\rangle = -\frac 12 | \Phi \rangle$, for the 
present model state $|\Phi\rangle$, the ground-state energy is now readily evaluated from 
Eq.\ (\ref{gsenergy}) as
\begin{equation}
{E_g} =  \langle \Phi | {\rm e}^{-\hat{S}}\hat{H}{\rm e}^{\hat{S}} | \Phi \rangle  ~~ \Rightarrow ~~ \frac {E_g}N = 
-\frac 14  - \frac {\lambda a}2 \; .
\label{ising_sub1_GSenergy_prelim}
\end{equation}
Furthermore the ground-state LSUB1 equation for the coefficient $a$ is given from 
Eq.\ (\ref{ket_state_eqn}) as follows,
\begin{equation}
\frac 1N \sum_{l=1}^N
\langle \Phi | \hat{s}_l^-  {\rm e}^{-\hat{S}}\hat{H}{\rm e}^{\hat{S}} | \Phi \rangle = 0  ~~ \Rightarrow ~~ 
a - \frac \lambda 2 + \frac {\lambda a^2}2 = 0 \; .
\label{ising_sub1_ket_1}
\end{equation}
This quadratic equation has the physical solution
\begin{equation}
a = \frac 1{\lambda} \left (-1 + \sqrt{\lambda^2 + 1 }  \right ) \; ,
\label{ising_sub1_ket_2}
\end{equation}
where we have discarded the (unphysical) solution with the negative sign of the 
square root since $a$ must be zero when $\lambda=0$, as noted above.
Using Eqs.\ (\ref{ising_sub1_GSenergy_prelim}) and (\ref{ising_sub1_ket_2}), 
the ground-state energy is thus given by
\begin{equation}
\frac {E_g}N = \frac 14  - \frac 12 \sqrt{\lambda^2 + 1 } \; ,
\label{ising_sub1_GSenergy}
\end{equation}
at the CCM LSUB1 level of approximation.
As tests of this equation, we note that $E_g/N = - 1/4$ when $\lambda=0$ and that $E_g/N \rightarrow - \lambda/2$ as $\lambda \rightarrow \infty$, which are the correct results in these two limiting cases. 
For comparison both with the exact result of 
Eq.\ (\ref{ising-exact-E}) and with those from higher-order CCM LSUB$n$ approximations
with $n>1$, the LSUB1 result of Eq.\ (\ref{ising_sub1_GSenergy}) is also shown in
Fig.~\ref{ti_energies}.

At the same CCM LSUB1 level of approximation the bra-state $\hat{\tilde S}$ operator is given by
\begin{equation}
\hat{\tilde S} = 1 + \tilde a \sum_{i=1}^{N} \hat{s}_i^- : ,
\label{bra_sub1}
\end{equation}
such that we may now evaluate the LSUB1 ground-state energy expectation value functional, $\bar H$, 
as
\begin{equation}
\bar H \equiv   \langle \tilde \Psi | \hat{H} | \Psi \rangle = \langle \Phi | \hat{\tilde S} {\rm e}^{-\hat{S}}\hat{H} {\rm e}^{\hat{S}} | \Phi \rangle ~~ \Rightarrow ~~ \frac 1N \bar H = 
-\frac 14  - \frac {\lambda a}2 + \tilde a \biggl ( a - \frac \lambda 2 + \frac {\lambda a^2}2  \biggr ) 
\; . 
\end{equation}
We see immediately that
\begin{equation}
\frac 1N 
\frac {\partial \bar H} {\partial \tilde a} = 0 ~~ \Rightarrow ~~ 
a - \frac \lambda 2 + \frac {\lambda a^2}2  = 0 
\; ,
\end{equation}
which simply gives us the LSUB1 ket-state equation of Eq.\ (\ref{ising_sub1_ket_1}) again, as required.  Furthermore, we see that the LSUB1 bra-state coefficient $\tilde a$ can similarly now also be obtained as follows,
\begin{equation}
\frac 1N 
\frac {\partial \bar H} {\partial a} = 0 ~~ \Rightarrow ~~ 
\tilde a - \frac \lambda 2 + \lambda a \tilde a  = 0 
~~ \Rightarrow ~~ \tilde a  = \frac {\lambda}{2 \sqrt{\lambda^2 + 1 } } \; .
\label{ising_sub1_bra}
\end{equation}
The magnetisation in the Ising direction, 
\begin{equation}
M^z = -\frac 1N \sum_{i=1}^{N}   \langle \tilde \Psi | \hat{s}_i^z | \Psi \rangle \; ,
\label{M1}
\end{equation}
can now easily be evaluated at the LSUB1 level of approximation as
\begin{equation}
M^z = \frac 12 - \tilde{a}a 
= \frac 1{2 \sqrt{\lambda^2 + 1 } }\; . 
\label{ising_sub1_Mz}
\end{equation}
The transverse magnetisation is given by
\begin{equation}
M^{\rm trans.} = \frac 1N \sum_{i=1}^{N}  \langle \tilde \Psi | \hat{s}_i^x | \Psi \rangle ~~ 
= \frac 1{2N} \sum_{i=1}^{N}   \langle \Psi | (\hat{s}_i^+ + \hat{s}_i^-) | \Psi \rangle \; ,
\label{M2}
\end{equation}
which is readily evaluated at the LSUB1 level of approximation as
\begin{equation}
M^{\rm trans.}  = \frac a2 +  \frac {\tilde a}2 - \frac {\tilde a a^2}2 \;.
\label{ising_sub1_Mtrans-prelim}
\end{equation}
By substituting the explicit solutions for $a$ and $\tilde{a}$ from 
Eqs.\ (\ref{ising_sub1_ket_2}) and (\ref{ising_sub1_bra}), respectively, into 
Eq.\ (\ref{ising_sub1_Mtrans-prelim}), we readily find the result
\begin{equation}
M^{\rm trans.} =  \frac{\lambda}{{2 \sqrt{\lambda^2 + 1 }}}\; .
\label{ising_sub1_Mtrans}
\end{equation}
Once again, for comparison both with the corresponding exact results of 
Eqs.\ (\ref{ising-exact-Mz}) and (\ref{ising-exact-Mtrans}) and with those 
from higher-order CCM LSUB$n$ approximations
with $n>1$, the LSUB1 results of Eqs.\ (\ref{ising_sub1_Mz})
and (\ref{ising_sub1_Mtrans}) are also shown in
Figs.~\ref{ti_M} and \ref{ti_MT}, respectively.

We now carry out a CCM LSUB1 calculation for the 1D transverse Ising model of 
Eq.\ (\ref{ising2}), i.e., after the unitary transformation involving the
rotation of the\
local spin axes. The Hamiltonian of Eq.\ (\ref{ising2}) now 
contains terms with both real and imaginary coefficients, while the model state 
remains one in which all spins point in the downwards $z$-direction. 
We wish to compare LSUB1 results for this new Hamiltonian to those for 
the ``unrotated'' case with Hamiltonian given by Eq.\ (\ref{ising1}). 
Indeed, we show explicitly that macroscopic quantities for this 
Hamiltonian do not change (and remain real), as they must, for the LSUB1
approximation even though the CCM correlation coefficients may now be 
complex-valued (i.e., contain real and imaginary components). 

The LSUB1 approximation is again given by Eq.\ (\ref{sub1}) and the 
similarity transformed spin operators are given as before in Eq.\ (\ref{sub1_expansions}). 
Thus we see that
\begin{equation}
{\rm e}^{-\hat{S}}\hat{H}{\rm e}^{\hat{S}} =
 -\sum_{i=1}^{N} (\hat{s}_{i}^z \hat{s}_{i+1}^z  +  a \hat{s}_{i}^z \hat{s}_{i+1}^+ +  
a \hat{s}_{i}^+ \hat{s}_{i+1}^z + a^2 \hat{s}_{i}^+ \hat{s}_{i+1}^+ ) 
-\frac {{\rm i} \lambda }2  \sum_{i=1}^{N} (-\hat{s}_i^+ + \hat{s}_i^- - 2a \hat{s}_i^z - a^2 \hat{s}_i^+ ) \;.
\label{ising_tilde_2}
\end{equation}
The ground-state energy equation is now given by
\begin{equation}
{E_g} = \langle \Phi | {\rm e}^{-\hat{S}}\hat{H}{\rm e}^{\hat{S}} | \Phi \rangle  ~~ \Rightarrow ~~ \frac {E_g}N = 
-\frac 14  - \frac {\lambda a {\rm i} }2 \; ,
\label{H3}
\end{equation}
and the ground-state energy LSUB1 equation is given by 
\begin{equation}
\frac 1N \sum_{l=1}^N
\langle \Phi | \hat{s}_l^-  {\rm e}^{-\hat{S}}\hat{H}{\rm e}^{\hat{S}} | \Phi \rangle = 0   ~~ \Rightarrow ~~ 
a + \frac {{\rm i} \lambda  } 2 + \frac {{\rm i} \lambda  a^2}2 = 0 
\; .
\end{equation}
Again, this quadratic equation has the physical solution given by
\begin{equation}
a = \frac {\rm i}{\lambda} \biggl (1 - \sqrt{\lambda^2 + 1 }  \biggr )\; ,
\label{temp1}
\end{equation}
where we have now discarded the (unphysical) solution with the positive sign of the 
square root since $a$ must be zero when $\lambda=0$, as noted previously.
Thus, we see that the new LSUB1 ket-state correlation coefficient $a$ is now
complex-valued (indeed, pure imaginary) for all values of $\lambda (> 0)$.   
Substitution of this value for $a$ from Eq.\ (\ref{temp1}) into Eq.\ (\ref{H3}) 
immediately yields that the ground-state energy at the CCM 
LSUB1 level of approximation is given by
\begin{equation}
\frac {E_g}N = 
\frac 14  - \frac 12 \sqrt{\lambda^2 + 1 } \; ,
\end{equation}
which is indeed seen to be identical to that given by Eq.\ (\ref{ising_sub1_GSenergy}).
The bra-state $S$ operator is again given by Eq.\ (\ref{bra_sub1}), 
such that we may explicitly evaluate the ground-state energy functional,
\begin{equation}
\bar H \equiv   \langle \tilde \Psi | \hat{H} | \Psi \rangle = \langle \Phi | \hat{\tilde S} {\rm e}^{-\hat{S}}\hat{H} {\rm e}^{\hat{S}} | \Phi \rangle ~~ \Rightarrow ~~ \frac 1N \bar H  = 
-\frac 14  - \frac {{\rm i} \lambda  a}2 + \tilde a \biggl ( a + \frac {{\rm i} \lambda } 2 + \frac 
{{\rm i} \lambda  a^2}2  \biggr ) 
\; . 
\end{equation}
We see immediately that
\begin{equation}
\frac 1N 
\frac {\partial \bar H} {\partial \tilde a} = 0 ~~ \Rightarrow ~~ 
a + \frac {{\rm i} \lambda } 2 + \frac {{\rm i} \lambda  a^2}2  = 0 
\; ,
\end{equation}
again as required. Furthermore, we see that
\begin{equation}
\frac 1N 
\frac {\partial \bar H} {\partial a} = 0 ~~ \Rightarrow ~~ 
\tilde a - \frac {{\rm i} \lambda } 2 + {\rm i} \lambda  a \tilde a  = 0 
~~ \Rightarrow ~~ \tilde a  = \frac {{\rm i} \lambda }{2 \sqrt{\lambda^2 + 1 } } \; .
\label{temp2}
\end{equation}

The magnetisation in the Ising direction is again given by Eq.\ (\ref{M1}), 
which at the LSUB1 level of approximation (now using 
Eqs. (\ref{temp1}) and (\ref{temp2})) is given by
\begin{equation}
M^z = \frac 12 - a \tilde a 
= \frac 1{2 \sqrt{\lambda^2 + 1 } }\; , 
\end{equation}
which is now explicitly real and in agreement with Eq.\ (\ref{ising_sub1_Mz}),
both as required.
The transverse magnetisation should now, of course, be evaluated with 
respect to the $y$-direction 
(in terms of spin coordinates after rotation of the local spin axes) and it is given by
\begin{equation}
M^{\rm trans.} = \frac 1N \sum_{i=1}^{N}   \langle \tilde \Psi | \hat{s}_i^y | \Psi \rangle ~~ 
= \frac {\rm i}{2N} \sum_{i=1}^{N}   \langle \Psi | (\hat{s}_i^- - \hat{s}_i^+) | \Psi \rangle \; ,
\label{M3}
\end{equation}
which at the LSUB1 level of approximation, as defined by Eqs.\ (\ref{sub1}) 
and (\ref{bra_sub1}),
is readily evaluated as
\begin{equation}
M^{\rm trans.}  = \frac {{\rm i} a}2 -  \frac {{\rm i} \tilde a}2 - \frac {{\rm i} \tilde a a^2}2  
\; .
\label{ising_sub1_Mtrans-prelim_2}
\end{equation}
Direct substitution into Eq.\ (\ref{ising_sub1_Mtrans-prelim_2}) with
the LSUB1 solutions for $a$ and $\tilde{a}$ from Eqs.\ (\ref{temp1}) 
and (\ref{temp2}), respectively, readily
yields the explicit expression,
\begin{equation}
M^{\rm trans.} = \frac \lambda {2 \sqrt{\lambda^2 + 1 } }\; . 
\end{equation}
which is again explicitly real and in agreement with Eq.\ (\ref{ising_sub1_Mtrans}),
both as required.

Thus, we have explicitly shown that the results for the ground-state energy per spin 
and the magnetisations in the Ising and transverse directions
at the LSUB1 level of approximation for the Hamiltonian of
Eq.\ (\ref{ising2}), which contains complex-valued coefficients, are not only real numbers, 
even though both the ket- and bra-states correlation coefficients are 
demonstrably complex-valued, but they are also identical to the corresponding results 
derived previously for the unitarily equivalent Hamiltonian of Eq.\ (\ref{ising1}). 

\section{Calculation of the Rotated Hamiltonian for Model State V}
\label{App-B}
We start from a Hamiltonian given by Eq.\ (\ref{heisenberg}), i.e., 
\begin{equation}
\hat{H} = \sum_{\langle i,j \rangle} \hat{{\bf s}}_i  \cdot  \hat{{\bf s}}_j - 
\lambda \sum_{i=1}^{N} \hat{s}_i^z 
=\sum_{\langle i,j \rangle}\bigl (\hat{s}_{i}^x \hat{s}_{j}^x +  
\hat{s}_{i}^y \hat{s}_{j}^y  + \hat{s}_{i}^z \hat{s}_{j}^z \bigr )
 - \lambda \sum_{i=1}^{N} \hat{s}_i^z \;,
\end{equation}
where the sum over the index $\langle i,j \rangle$ indicate a sum over all nearest-neighbour 
pairs on the triangular lattice, with each pair being counted once and once only. 
We now carry out the first of a number of unitary transformations of the local spin axes 
given by,
\begin{equation}
\hat{s}_i^x \rightarrow \hat{s}_i^x\; ; ~~ \hat{s}_i^y \rightarrow -\hat{s}_i^z\; ; 
~~ \hat{s}_i^z \rightarrow \hat{s}_i^y \;,
\end{equation}
where $i$ runs over all lattice sites on the triangular lattice. The Hamiltonian is now given by
\begin{equation}
\hat{H}= \sum_{\langle i,j \rangle} \bigl (\hat{s}_{i}^x \hat{s}_{j}^x +  
\hat{s}_{i}^y \hat{s}_{j}^y  + \hat{s}_{i}^z \hat{s}_{j}^z\bigr )
 - \lambda \sum_{i=1}^{N} \hat{s}_i^y \; ,
 \label{state-V_Hamiltonian-after-rot1}
\end{equation}

This first transformation represents a rotation by $90^{\circ}$ 
about the $x$-axis. We do this so that the subsequent rotations of the model state 
are both slightly easier to formulate and follow in direct analogy to earlier work 
on the coplanar states for this model.  In particular, the next set of rotations 
of spins in the $xz$-plane follow exactly that set out in Ref.\ \cite{ccm5} for 
the triangular lattice with zero external field.  We first define three 
interpenetrating  sublattices $(A, B, C)$ for the triangular lattice,
such that each elementary triangular plaquette formed from nearest-neighbour 
sites of the original lattice contains one site from each of the three
sublattices.  Hence, we start now with the 
the usual $120^\circ$ three-sublattice N\'{e}el state, as shown for state I in
Fig.~\ref{model_states}, with $\alpha = 30^\circ$.  The spins all lie in the 
$xz$-plane, with spins on sublattice $C$ aligned along the negative $z$-axis 
and those on sublattices $A$ and $B$ oriented respectively at angles $-120^\circ$ and 
$+120^\circ$ with respect to those on sublattice $C$.  We now perform the necessary passive
rotations so that all spins point downwards (i.e., in the negative $z$-direction).
Thus, we rotate the axes for spins at sites $i_A$ on 
the $A$-sublattice by $+120^\circ$ about the $y$-axis via:
\begin{equation}
\hat{s}_{i_A}^x \rightarrow -\frac 12 \hat{s}_{i_A}^x - \frac {\sqrt{3}}2 \hat{s}_{i_A}^z\;; \quad
\hat{s}_{i_A}^y \rightarrow \hat{s}_{i_A}^y \;; \quad
\hat{s}_{i_A}^z \rightarrow \frac {\sqrt{3}}2 \hat{s}_{i_A}^x -\frac 12 \hat{s}_{i_A}^z\;. 
\label{rotation_A-sublatt}
\end{equation}
Simultaneously, we rotate the axes for spins at sites $i_B$ on 
the $B$-sublattice by $-120^\circ$ about the $y$-axis via:
\begin{equation}
\hat{s}_{i_B}^x \rightarrow -\frac 12 \hat{s}_{i_B}^x + \frac {\sqrt{3}}2  \hat{s}_{i_B}^z \;; \quad
\hat{s}_{i_B}^y \rightarrow \hat{s}_{i_B}^y\;; \quad
\hat{s}_{i_B}^z \rightarrow - \frac {\sqrt{3}}2 \hat{s}_{i_B}^x -\frac 12 \hat{s}_{i_B}^z\;.
\label{rotation_B-sublatt}
\end{equation}
Furthermore, in this step, we do not rotate the axes for spins at sites $i_C$ on the $C$-sublattice, since they are already pointing in the downwards direction: 
\begin{equation}
\hat{s}_{i_C}^x \rightarrow \hat{s}_{i_C}^x  \;; \quad
\hat{s}_{i_C}^y \rightarrow \hat{s}_{i_C}^y \;; \quad
\hat{s}_{i_C}^z \rightarrow \hat{s}_{i_C}^z\;.
\label{rotation_C-sublatt}
\end{equation}

We may now use Eqs.\ (\ref{rotation_A-sublatt})--(\ref{rotation_C-sublatt}) to rewrite the nearest-neighbour part of the Hamiltonian of Eq.\ (\ref{state-V_Hamiltonian-after-rot1}) that connects sites $i_B$ on the $B$-sublattice and $j_C$ on the $C$-sublattice in the direction going {\it from} site $i_B$ {\it to} site $j_C$ in the (unitarily equivalent) rotated  form
\begin{eqnarray}
\hat{H}_{i_B \rightarrow j_C}&\equiv& \sum_{\langle i_B \rightarrow j_C \rangle}\bigl \{ \hat{s}_{i_B}^x \hat{s}_{j_C}^x +  \hat{s}_{i_B}^y \hat{s}_{j_C}^y  + \hat{s}_{i_B}^z \hat{s}_{j_C}^z \bigr \}   \nonumber \\
&=& \sum_{\langle i_B \rightarrow j_C \rangle} \biggl \{ \biggl (-\frac 12 \hat{s}_{i_B}^x +\frac {\sqrt{3}}2  \hat{s}_{i_B}^z \biggr ) \hat{s}_{j_C}^x  +  \hat{s}_{i_B}^y \hat{s}_{j_C}^y  
+ \biggl (-\frac {\sqrt{3}}2  \hat{s}_{i_B}^x - \frac 12 \hat{s}_{i_B}^z \biggr ) \hat{s}_{j_C}^z \biggr \}  
\nonumber \\
&=& \sum_{\langle i_B \rightarrow j_C \rangle}\biggl \{ -\frac 12 \hat{s}_{i_B}^x \hat{s}_{j_C}^x +\frac {\sqrt{3}}2 
\bigl (\hat{s}_{i_B}^z \hat{s}_{j_C}^x   -  \hat{s}_{i_B}^x \hat{s}_{j_C}^z \bigr ) +   \hat{s}_{i_B}^y \hat{s}_{j_C}^y  -  \frac 12 \hat{s}_{i_B}^z \hat{s}_{j_C}^z \biggr \}  \;.
\label{state-V_NN-Hamiltonian-BtoC}
\end{eqnarray}
We may similarly use Eqs.\ (\ref{rotation_A-sublatt})--(\ref{rotation_C-sublatt}) to rewrite the nearest-neighbour part of the Hamiltonian of Eq.\ (\ref{state-V_Hamiltonian-after-rot1}) that connects sites $i_C$ on the $C$-sublattice and $j_A$ on the $A$-sublattice in the direction going {\it from} site $i_C$ {\it to} site $j_A$ in the rotated  form
\begin{eqnarray}
\hat{H}_{i_C \rightarrow j_A}&\equiv& \sum_{\langle i_C \rightarrow j_A \rangle}\bigl \{ \hat{s}_{i_C}^x \hat{s}_{j_A}^x +  \hat{s}_{i_C}^y \hat{s}_{j_A}^y  + \hat{s}_{i_C}^z \hat{s}_{j_A}^z \bigr \}  \nonumber \\
&=& \sum_{\langle i_C \rightarrow j_A \rangle}\biggl \{ \hat{s}_{i_C}^x \biggl (-\frac 12 \hat{s}_{j_A}^x - \frac {\sqrt{3}}2 \hat{s}_{j_A}^z \biggr) + \hat{s}_{i_C}^y \hat{s}_{j_A}^y 
+ \hat{s}_{j_A}^z \biggl ( \frac {\sqrt{3}}2 \hat{s}_{j_A}^x -\frac 12 \hat{s}_{j_A}^z \biggr ) \biggr \}  
\nonumber \\
&=& \sum_{\langle i_C \rightarrow j_A \rangle}\biggl \{ -\frac 12 \hat{s}_{i_C}^x \hat{s}_{j_A}^x +\frac {\sqrt{3}}2  \bigl (\hat{s}_{i_C}^z \hat{s}_{j_A}^x   -  \hat{s}_{i_C}^x \hat{s}_{j_A}^z \bigr ) +   \hat{s}_{i_C}^y \hat{s}_{j_A}^y  -  \frac 12 \hat{s}_{i_C}^z \hat{s}_{j_A}^z \biggr \}  \;.
\label{state-V_NN-Hamiltonian-CtoA}
\end{eqnarray}
Lastly, we again use Eqs.\ (\ref{rotation_A-sublatt})--(\ref{rotation_C-sublatt}) to rewrite the nearest-neighbour part of the Hamiltonian of Eq.\ (\ref{state-V_Hamiltonian-after-rot1}) that connects sites $i_A$ on the $A$-sublattice and $j_B$ on the $B$-sublattice in the direction going {\it from} site $i_A$ {\it to} site $j_B$ in the rotated  form
\begin{eqnarray}
\hat{H}_{i_A \rightarrow j_B}&\equiv& \sum_{\langle i_A \rightarrow j_B \rangle}\bigl \{ \hat{s}_{i_A}^x \hat{s}_{j_B}^x +  \hat{s}_{i_A}^y \hat{s}_{j_B}^y  + \hat{s}_{i_A}^z \hat{s}_{j_B}^z \bigr \}  \nonumber \\
&=& \sum_{\langle i_A \rightarrow j_B \rangle}\ \biggl \{ \biggl(-\frac 12 \hat{s}_{i_A}^x - \frac {\sqrt{3}}2 \hat{s}_{i_A}^z \biggr ) 
\biggl (-\frac 12 \hat{s}_{j_B}^x + \frac {\sqrt{3}}2 \hat{s}_{j_B}^z\biggr ) +  \hat{s}_{i_A}^y \hat{s}_{j_B}^y 
\nonumber \\ & &\quad \quad \quad ~
+\biggl ( \frac {\sqrt{3}}2 \hat{s}_{i_A}^x -\frac 12 \hat{s}_{i_A}^z\biggr ) \biggl (-\frac {\sqrt{3}}2 \hat{s}_{j_B}^x -\frac 12 \hat{s}_{j_B}^z \biggr ) \biggr \}  \nonumber \\
&=& \sum_{\langle i_A \rightarrow j_B \rangle}\biggl \{ -\frac 12 \hat{s}_{i_A}^x \hat{s}_{j_B}^x +\frac {\sqrt{3}}2 \bigl (\hat{s}_{i_A}^z \hat{s}_{j_B}^x   -  \hat{s}_{i_A}^x \hat{s}_{j_B}^z \bigr ) +   \hat{s}_{i_A}^y \hat{s}_{j_B}^y  -  \frac 12 \hat{s}_{i_A}^z \hat{s}_{j_B}^z \biggr \} \;.
\label{state-V_NN-Hamiltonian-AtoB}
\end{eqnarray}
By making use of 	Eqs.\ (\ref{state-V_NN-Hamiltonian-BtoC})--(\ref{state-V_NN-Hamiltonian-AtoB}), the Hamiltonian of Eq.\ (\ref{state-V_Hamiltonian-after-rot1}) may now be written as
\begin{equation}
\begin{split}
\hat{H} =  \sum_{\langle i_{B,C,A} \rightarrow j_{C,A,B} \rangle}\biggl \{ &-\frac 12 \hat{s}_{i_{B,C,A}}^x \hat{s}_{j_{C,A,B}}^x +\frac {\sqrt{3}}2  \bigl (\hat{s}_{i_{B,C,A}}^z \hat{s}_{j_{C,A,B}}^x   -  \hat{s}_{i_{B,C,A}}^x \hat{s}_{j_{C,A,B}}^z \bigr ) \\   
&+ \hat{s}_{i_{B,C,A}}^y \hat{s}_{j_{C,A,B}}^y   
-  \frac 12 \hat{s}_{i_{B,C,A}}^z \hat{s}_{j_{C,A,B}}^z \biggr \}  - 
\lambda \sum_{i=1}^N \hat{s}_i^y \;,
\label{state-V_Hamiltonian-after-rot2}
\end{split}
\end{equation}
after the second set of rotations of the local spin axes have been made, and where 
the sum over $\langle i_{B,C,A} \rightarrow j_{C,A,B} \rangle$ represent a 
shorthand notation to include the  three sorts of ``directed'' nearest-neighbour 
bonds on each basic triangular plaquette of side $a$ on the triangular lattice, 
which join sites $i_B$ and $j_C$ going from the 
$B$-sublattice to the $C$-sublattice, sites $i_C$ and $j_A$ going from 
the $C$-sublattice to the $A$-sublattice, and sites $i_A$ and $j_B$ going 
from the $A$-sublattice to the $B$-sublattice 
(in those directions only and not reversed)..

The effects of an external field for this model state can then be included straightforwardly by a final rotation in the $yz$-plane (i.e., about the $x$-axis)
by an angle of $\theta$ for {\it all} spins (i.e., on all sublattices). This is
performed via the following transformation:
\begin{equation}
\hat{s}_{i}^x \rightarrow \hat{s}_{i}^x\;; \quad
\hat{s}_{i}^y \rightarrow \cos \theta \, \hat{s}_{i}^y  - \sin \theta \, \hat{s}_{i}^z\;; \quad
\hat{s}_{i}^z \rightarrow \sin \theta \, \hat{s}_{i}^y  + \cos \theta \, \hat{s}_{i}^z \;.
\end{equation}
The Hamiltonian of Eq.\ (\ref{state-V_Hamiltonian-after-rot2}) is thus now 
\begin{eqnarray}
\hat{H}&=&  \sum_{\langle i_{B,C,A} \rightarrow j_{C,A,B} \rangle} \biggl \{ \frac {\sqrt{3}}2 \Bigl [ \bigl ( \sin \theta\, \hat{s}_{i_{B,C,A}}^y  + \cos \theta \, \hat{s}_{i_{B,C,A}}^z \bigr ) \hat{s}_{j_{C,A,B}}^x   -  \hat{s}_{i_{B,C,A}}^x \bigl (\sin \theta \, \hat{s}_{j_{C,A,B}}^y  + \cos \theta \, \hat{s}_{j_{C,A,B}}^z \bigr ) \Bigr ]
 \nonumber \\ & &\quad \quad \quad \quad \quad \quad
-\frac 12 \hat{s}_{i_{B,C,A}}^x \hat{s}_{j_{C,A,B}}^x +  \bigl (\cos \theta \, \hat{s}_{i_{B,C,A}}^y  - \sin \theta \, \hat{s}_{i_{B,C,A}}^z \bigr ) \bigl (\cos \theta \, \hat{s}_{j_{C,A,B}}^y  - \sin \theta \, \hat{s}_{j_{C,A,B}}^z \bigr )  
 \nonumber \\ & &\quad \quad \quad \quad \quad \quad
-  \frac 12 \bigl (\sin \theta \, \hat{s}_{i_{B,C,A}}^y  + \cos \theta\, \hat{s}_{i_{B,C,A}}^z \bigr ) \bigl (\sin \theta \, \hat{s}_{j_{C,A,B}}^y  + \cos \theta \, \hat{s}_{j_{C,A,B}}^z\bigr ) \biggr \}   \nonumber \\ & & \quad
- \lambda \sum_{i=1}^{N}\bigl [ \cos \theta \, \hat{s}_i^y  - \sin \theta \, \hat{s}_i^z \bigr ]
 \nonumber \\ 
 &=& \sum_{\langle i_{B,C,A} \rightarrow j_{C,A,B} \rangle}\biggl \{  
\biggl (-\frac 12 \cos^2 \theta+ \sin^2 \theta \biggr) \hat{s}_{i_{B,C,A}}^z \hat{s}_{j_{C,A,B}}^z  -\frac{1}{2} \hat{s}_{i_{B,C,A}}^x \hat{s}_{j_{C,A,B}}^x 
\nonumber \\ & &\quad \quad \quad \quad \quad \quad
   +\biggl (-\frac 12 \sin^2 \theta+ \cos^2 \theta \biggr) \hat{s}_{i_{B,C,A}}^y \hat{s}_{j_{C,A,B}}^y   \nonumber \\ & &\quad \quad \quad \quad \quad \quad 
 +  \frac {\sqrt{3}}{2} \cos \theta \, \bigl (\hat{s}_{i_{B,C,A}}^z \hat{s}_{j_{C,A,B}}^x   -  \hat{s}_{i_{B,C,A}}^x \hat{s}_{j_{C,A,B}}^z \bigr )
\nonumber \\ & &\quad \quad \quad \quad \quad \quad 
+ \frac {\sqrt{3}}{2} \sin \theta \, \bigl (\hat{s}_{i_{B,C,A}}^y \hat{s}_{j_{C,A,B}}^x   -  \hat{s}_{i_{B,C,A}}^x \hat{s}_{j_{C,A,B}}^y \bigr )
\nonumber \\ & &\quad \quad \quad \quad \quad \quad
  - \frac{3}{2} \sin \theta \cos \theta \,  \bigl (\hat{s}_{i_{B,C,A}}^y \hat{s}_{j_{C,A,B}}^z   +  \hat{s}_{i_{B,C,A}}^z \hat{s}_{j_{C,A,B}}^y \bigr ) \biggr \}
\nonumber \\ & &\quad
- \lambda \sum_{i=1}^{N}\bigl [ \cos \theta \, \hat{s}_i^y  - \sin \theta \, \hat{s}_i^z \bigr ] \;.
    \label{rotH4-interim} 
\end{eqnarray}
As shown in Fig. \ref{angle}, we obtain a minimal energy solution for 
$\theta=0$ (i.e., where all spins lie in the $xz$-plane) when the external field 
$\lambda$ is zero, whereas we obtain a minimal energy solution for $\theta=\pi/2$ 
(i.e., where all spins point along the $y$-axis) when $\lambda$ reaches the 
saturation field, $\lambda_s=4.5$, for the spin-$\frac{1}{2}$ model. 
Equation (\ref{rotH4-interim}) may then be rewritten in our final form,
\begin{equation}
\begin{split}
\hat{H} &=
\sum_{\langle i_{B,C,A} \rightarrow j_{C,A,B} \rangle} \biggl \{  
\left(\sin^2 \theta - \frac{1}{2} \cos^2 \theta \right) \hat{s}_{i_{B,C,A}}^z \hat{s}_{j_{C,A,B}}^z \\ 
&  \qquad \qquad \qquad 
+ \frac{1}{4} \left( \frac{1}{2} \sin^2 \theta - \cos^2 \theta -  \frac{1}{2} \right) \bigl (\hat{s}_{i_{B,C,A}}^+ \hat{s}_{j_{C,A,B}}^+ + \hat{s}_{i_{B,C,A}}^- \hat{s}_{j_{C,A,B}}^- \bigr )
\\ & \qquad \qquad \qquad 
+ \frac{1}{4} \left(\cos^2 \theta -  \frac{1}{2} \sin^2 \theta - \frac{1}{2} \right) \bigl (\hat{s}_{i_{B,C,A}}^+ \hat{s}_{j_{C,A,B}}^- + \hat{s}_{i_{B,C,A}}^- \hat{s}_{j_{C,A,B}}^+ \bigr ) \\ &  \qquad \qquad \qquad 
+ \frac{\sqrt{3}}{4}  \cos \theta \, \Bigl [ \hat{s}_{i_{B,C,A}}^z \bigl (\hat{s}_{j_{C,A,B}}^+ + \hat{s}_{j_{C,A,B}}^- \bigr ) - \bigl ( \hat{s}_{i_{B,C,A}}^+ + \hat{s}_{i_{B,C,A}}^- \bigr ) \hat{s}_{j_{C,A,B}}^z \Bigr ]
\biggr \} \\ 
 &  \quad \quad
 + \lambda \sum_{i=1}^{N}  \sin \theta \, \hat{s}_{i}^z  \\ 
 & \quad \quad
 + {\rm i}   \sum_{\langle i_{B,C,A} \rightarrow j_{C,A,B} \rangle} \biggl \{  
 \frac{\sqrt{3}}{4}  \sin \theta \, \bigl ( \hat{s}_{i_{B,C,A}}^- \hat{s}_{j_{C,A,B}}^+ - \hat{s}_{i_{B,C,A}}^+ \hat{s}_{j_{C,A,B}}^-\bigr ) \\ 
 &  \qquad \qquad \qquad  \!
+ \frac{3}{4} \sin \theta  \cos\theta \, \Bigl [\hat{s}_{i_{B,C,A}}^z \bigl ( \hat{s}_{j_{C,A,B}}^+ - \hat{s}_{j_{C,A,B}}^- \bigr ) + \bigl ( \hat{s}_{i_{B,C,A}}^+ - \hat{s}_{i_{B,C,A}}^- \bigr ) \hat{s}_{j_{C,A,B}}^z\Bigr ]
 \biggr \}  \\
& \quad \quad
+ {\rm i}\,   \frac{\lambda}{2} \sum_{i=1}^{N} \cos \theta \, \bigl ( \hat{s}_{i}^+ - \hat{s}_{i}^- \bigr ) \;.
    \label{rotH4} 
\end{split}    
\end{equation}

%\pagebreak

\end{document}